\documentclass[aps,prl,floatfix,superscriptaddress,nofootinbib,
amsmath,amssymb,twocolumn,preprintnumbers]{revtex4-2}
\pdfoutput=1
\usepackage[dvipsnames]{xcolor}
\usepackage{graphicx}
\usepackage[colorlinks=true,citecolor=blue,linkcolor=blue,breaklinks=true]{hyperref}
\usepackage{soul}
\usepackage{mathtools}
\usepackage{multirow}
\usepackage{gensymb}
\usepackage{orcidlink}

\newcommand{\ket}[1]{\lvert#1\rangle}
\newcommand{\bra}[1]{\langle#1\rvert}

\newcommand{\tauDM}{\tau_{\mathrm{DM}}}
\newcommand{\E}{\mathbb E}

\let\oldsection\section
\renewcommand{\section}[1]{%
  \par\addvspace{0.5\baselineskip}
  \noindent\textit{\textbf{#1.---\hspace{-5pt}}}
}

\begin{document}
\title{Quantum Computing Enhanced Search for Wavelike Dark Matter
}

\author{Yunjia Bao%
\orcidlink{https://orcid.org/0000-0002-2622-0939}}
\email{yunjia.bao@uchicago.edu}
\affiliation{Department of Physics, Enrico Fermi Institute, Kavli Institute for Cosmological Physics, \\ University of Chicago, Chicago, IL 60637, USA}
\affiliation{Leinweber Institute for Theoretical Physics, University of Chicago, Chicago, IL 60637, USA}

\author{Yu Cheng%
\orcidlink{0000-0002-4822-3890}}
\email{chengyu059@uchicago.edu}
\affiliation{Department of Physics, Enrico Fermi Institute, Kavli Institute for Cosmological Physics, \\ University of Chicago, Chicago, IL 60637, USA}
\affiliation{Leinweber Institute for Theoretical Physics, University of Chicago, Chicago, IL 60637, USA}
\affiliation{Department of Physics, Korea Advanced Institute of Science and Technology (KAIST), Daejeon 34141, Korea}

\author{Liang Jiang%
\orcidlink{https://orcid.org/0000-0002-0000-9342}}
\email{liang.jiang@uchicago.edu}
\affiliation{Pritzker School of Molecular Engineering, University of Chicago, Chicago, Illinois 60637, USA}
\affiliation{Leinweber Institute for Theoretical Physics, University of Chicago, Chicago, IL 60637, USA}

\author{Lian-Tao Wang%
\orcidlink{https://orcid.org/0000-0001-6826-5142}}
\email{liantaow@uchicago.edu}
\affiliation{Department of Physics, Enrico Fermi Institute, Kavli Institute for Cosmological Physics, \\ University of Chicago, Chicago, IL 60637, USA}
\affiliation{Leinweber Institute for Theoretical Physics, University of Chicago, Chicago, IL 60637, USA}

\begin{abstract}
Identifying a weak, stochastic signal hidden within a broad frequency band is a central challenge in the search for wavelike dark matter.
We explore quantum computing enhanced sensing based on the Grover search algorithm to address this problem and find that a long integration time stabilizes the accumulated sensing phase, 
reducing quantum signal processing (QSP) errors induced by dark matter fluctuations.
We further propose alternating the signs of the QSP angles between oracle calls to suppress the coherent accumulation of the intrinsic QSP approximation errors during the search. 
Combining these strategies in a transmon-based benchmark, 
we obtain a projected sensitivity to axion dark matter beyond existing astrophysical limits.
This opens access to substantial unexplored parameter space across the
$10$–$100\,\mu\mathrm{eV}$ mass range and offers an accelerated discovery channel for wavelike dark matter.
With $10^6$ entangled sensing qubits, 
the projected sensitivity reaches the QCD axion band. 
Although illustrated with transmons, the protocol can be adapted to other experimental platforms for wavelike dark matter detection.
\end{abstract}

\maketitle

\section{Introduction}
Dark matter (DM) accounts for more than 80\% of the matter in the Universe, yet its properties, in particular its mass, remain unknown. 
A class of well-motivated DM models is the ultralight DM, 
such as axions, axion-like particles, and dark photons, which can be approximated as classical waves~\cite{Bao:2025nsd, Bao:2026prd} with a typical frequency set by their mass. 
Quantum sensors can probe such wavelike DM through spin precession~\cite{Graham:2013gfa,Budker:2013hfa,Graham:2017ivz,Abel:2017rtm,Wu:2019exd,Garcon:2019inh,Roussy:2020ily,Aybas:2021nvn,Wei:2023rzs,Jiang:2021dby,Bloch:2021vnn,Bloch:2022kjm,Arza:2025kuh,Zhao:2025kdw,Rigoulet:2026eai}, 
shifts atomic transition frequencies~\cite{Arvanitaki:2014faa,VanTilburg:2015oza,Hees:2016gop,Kennedy:2020bac,Filzinger:2023zrs,Arakawa:2026mls}, 
or induced electric field in the cavity that excites superconducting qubits~\cite{Chen:2022quj,Chen:2024aya,Kang:2025kaf}.
Squeezing and entanglement can further improve sensitivity or signal rate~\cite{Zheng:2016qjv,Malnou:2018dxn,HAYSTAC:2020kwv,Wurtz:2021cnm,Brady:2022bus,Chen:2023swh,Ito:2023zhp,Zaheer:2023ulf,HAYSTAC:2023cam,HAYSTAC:2024jch,Sichanugrist:2024wfk,Shi:2024qhd,He:2025ovo,Zheng:2025qgv,Freiman:2025tse}.
Collective measurements~\cite{Chen:2025tgj} and quantum-error-correction-inspired protocols~\cite{Fukuda:2025afi} have also been proposed to suppress
sensor backgrounds,  
while spatially separated sensors can extract directional information of the wavelike DM~\cite{Fukuda:2025zcf}.
However, the bottleneck is not only the feeble signal but also the wide frequency band to search.
Theoretical models motivate a broad mass range for wavelike DM. 
For example, QCD axion production through vacuum or kinetic misalignment~\cite{Marsh:2015xka, Co:2019jts, OHare:2024nmr} can reproduce the observed DM abundance at masses from $\sim 10^{-11}\,\text{eV}$ to $\sim 10^{-1}\,\text{eV}$, with $10$ to $100\,\mu\text{eV}$ as a natural range, and axion-like particles may span an even larger mass range~\cite{Arias:2012az, Svrcek:2006yi, Arvanitaki:2009fg}, not even mentioning other wavelike DM candidates. 
With the DM mass unknown, 
one might ask, similar to the sensitivity enhancement using quantum sensing techniques, whether we can take quantum advantage to search a wide frequency band.

Quantum search offers a way to expedite this scan. 
While a classical strategy requires approximately $\mathcal{O}(N)$ queries to search among $N$ candidates, Grover’s algorithm~\cite{Grover:1996rk,Grover:1997fa} needs only $\mathcal{O}(\sqrt{N})$ queries to identify a marked state. 
A recent proposal~\cite{Allen:2025hdx} applies this principle to search for the frequency of a monochromatic field with a fixed but unknown frequency and phase.
In their scheme, the index register qubits set the energy splitting of the sensing qubit.
The sensor accumulates a signal-induced phase during each elementary sensing unit (ESU), and
quantum signal processing (QSP) 
maps a prescribed range of these phases to an approximately constant sign on the sensor state, thereby realizing the oracle~\cite{Low:2016sck,Martyn:2021eaf}.
Then, the Grover protocol expedites the signal identification, which can be confirmed by a dedicated measurement at that frequency.
Quantum computing enhanced sensing has recently been demonstrated experimentally in superconducting transmon cavity systems~\cite{Xu:2026eys,Ouellet:2026bvs}. 
In these implementations, cavity Fock states label candidate frequencies, and the unknown signal drives frequency-selective transmon rotations that generate the phase oracle.

However, the random amplitudes and initial phases produce a fluctuating waveform of DM~\cite{Foster:2017hbq,Centers:2019dyn,Cheong:2024ose} that hinders the original search protocol.
When an ESU's sensing time is comparable with DM coherence time, 
the waveform varies within the sensing interval, and each ESU  sees a different, essentially random DM intensity.
Consequently, the phase imprinted on the sensor fluctuates.
These variations change the interference within QSP. 
Thus, no QSP sequence can turn such fluctuating phase into a reliable constant sign.
Applying quantum search to DM therefore raises three related questions, 
can the fluctuating field generate a sufficiently reproducible oracle, 
how much physical sensing time does
this require, 
and how weak can the field be while a
single search still succeeds with a stated probability?

This paper answers these three questions for wavelike DM.
We address the first question by accumulating the off-resonant AC Stark phase over a long sensing time $T$, over which many weakly correlated fluctuations average out so that the relative error in the sensed phase is suppressed. 
At a fixed mean phase, a longer $T$ can also sense a weaker DM signal.
We then propagate the remaining phase fluctuations through QSP and the full register–sensor evolution to find the minimum detectable field at 95\% expected success probability per search.

\section{Stochastic DM wave and sensor response}
For wavelike DM with mass $m_a$ and local energy density $\rho_{\rm DM}$, the DM field at the sensor can be modeled as a 
linear combination of stochastic waves \cite{Foster:2017hbq},
\begin{equation}
 a_\xi(t)
 =
 \frac{\sqrt{\rho_{\rm DM}}}{m_a}
 \sum_j A_j\sqrt{f(v_j)\Delta v}\,
 \cos[(m_a+q_j)t+\varphi_j].
 \label{eq:stochasticfield}
\end{equation}
where $f(v)$ is the normalized DM speed distribution, 
$v_j$ is a representative speed in the $j$-th bin of width $\Delta v$, 
and $q_j=m_a v_j^2/2$ is the corresponding kinetic energy.
The amplitudes $A_j$ follow a Rayleigh distribution $p(A_j)=A_j e^{-A_j^2/2}$, 
while the phases $\varphi_j$ are uniform on $[0,2\pi)$, independently across modes. 
Once the initial configuration $\xi = \{A_j,\varphi_j\}$ is chosen, 
these coefficients determine the entire waveform. 
We adopt a truncated Standard Halo Model for the DM's speed distribution with characteristic speed $v_0 = 220\,{\rm km} \rm{s}^{-1}$. See SM Sec.~\ref{sm:field}.

Consider a sensing qubit $S$ with transition frequency $\omega_S$. 
Its response to wavelike DM depends on both the microscopic DM interaction and the properties of the sensor. 
In the sensor's energy basis, a diagonal interaction changes the sensor's level separation, while an off-diagonal interaction can transfer population between levels. 
Here, we focus on the latter case:
\begin{equation}
 H_0=\frac{\omega_S}{2}Z_S,\qquad
 H_{\rm sig}^{(\xi)}(t)=B_\xi(t)X_S.
 \label{eq:DMsensorresponse}
\end{equation}
where $X_S$ and $Z_S$ are the Pauli operators, and $B_\xi$ denotes the dimension-one DM-sensor coupling, which can be parametrized as
\begin{equation}
 B_\xi(t)=\frac{m_a B_{\rm rms}}{\sqrt{\rho_{\rm DM}}}a_\xi(t),
 \qquad \E [B_\xi(t)^2]=B_{\rm rms}^2.
 \label{eq:driverms}
\end{equation}
in which $\E$ denotes the ensemble average over the configurations $\xi = \{A_j,\varphi_j\}$, and $B_{\rm rms}$ is the root-mean-square (RMS) coupling strength. 
As a benchmark, $B_{\rm rms}$ is related to the axion-photon coupling as shown in Eq.~\eqref{eq:transmonrms}, but we keep it general as our quantum-computing-enhanced protocol applies to generic axion-sensor couplings.

To search the DM's frequency coherently, an index register $Q$ is coupled to the sensor such that different register states set the sensor's splittings during each sensing unitary called the elementary sensing unit (ESU). 
Its $N$ active states, $|k\rangle_Q$ with $k=1,\ldots,N$, label the frequency bins and set the sensor energy levels to $\pm\omega_k/2$. 
During an ESU starting at $t_0$ with a sensing time $T$, we modulate the signal $B(t)$ to $\chi(t-t_0)B(t)$ where $\chi$ is a real, dimensionless, smooth function satisfying $0\leq\chi\leq1$ and $\chi(0)=\chi(T)=0$. 
This can be realized, for instance, by a partial dynamical decoupling protocol.
The full Hamiltonian is
\begin{align}
 H_{\rm ESU}^{(\xi)}(t)
 &= \sum_{k=1}^N 
 \ket{k} \bra{k}_Q 
 \otimes H_k^{(\xi)}(t),\nonumber\\
 H_k^{(\xi)}(t)
 &= \frac{\omega_k}{2}Z_S
 + \chi(t-t_0) B_\xi(t) X_S.
 \label{eq:sensor}
\end{align}
To separate the oscillating DM drive from the sensor's free evolution, 
we work in the frame rotating at $m_a$, transforming the sensor state by $e^{i m_a t Z_S/2}$.
Applying the rotating-wave approximation (RWA), we obtain
\begin{align}
 \widetilde H_k^{(\xi)}(t)
 ={}&\frac{\delta_k}{2} Z_S
 + \frac{B_{\rm rms} \chi(t-t_0)}{2}
 \sum_j A_j\sqrt{f(v_j)
 \Delta v}\nonumber\\
 &\times 
   \left[e^{-i(q_jt+\varphi_j)}\sigma_+
   +e^{i(q_jt+\varphi_j)}\sigma_-\right].
 \label{eq:rotatingH}
\end{align}
Here $\sigma_\pm = (X_S \pm i Y_S)/2$ are the ladder operator and $\delta_k = \omega_k - m_a$ is the detuning. 

The above equation determines the evolution of the sensor state. Its evolution matrix is therefore $\mathcal T\exp[-i\int_{t_0}^{t_0+T}{\rm d} t\, \widetilde H_k^{(\xi)}(t)]$, 
where $\mathcal T$ denotes the time ordering. 
This matrix includes both the response to DM and the free evolution.
Removing the latter in the
interaction picture of $\delta_k Z_S/2$ gives the ESU evolution matrix
\begin{equation}
 \begin{aligned}
 U_k^{(\xi)}(t_0,T) = {}&e^{i\delta_k (t_0+T) Z_S/2}\,\\
 &\times\mathcal T e^{-i\int_{t_0}^{t_0+T}\widetilde H_k^{(\xi)}(t){\rm d} t}
 e^{-i\delta_kt_0Z_S/2}.
 \end{aligned}
 \label{eq:ESU}
\end{equation}
This operation reduces
to the identity $I_S$ without a signal and otherwise captures the signal-induced phase and population transfer.

\begin{figure*}[!t]
 \centering
 \includegraphics[width=0.90\textwidth]{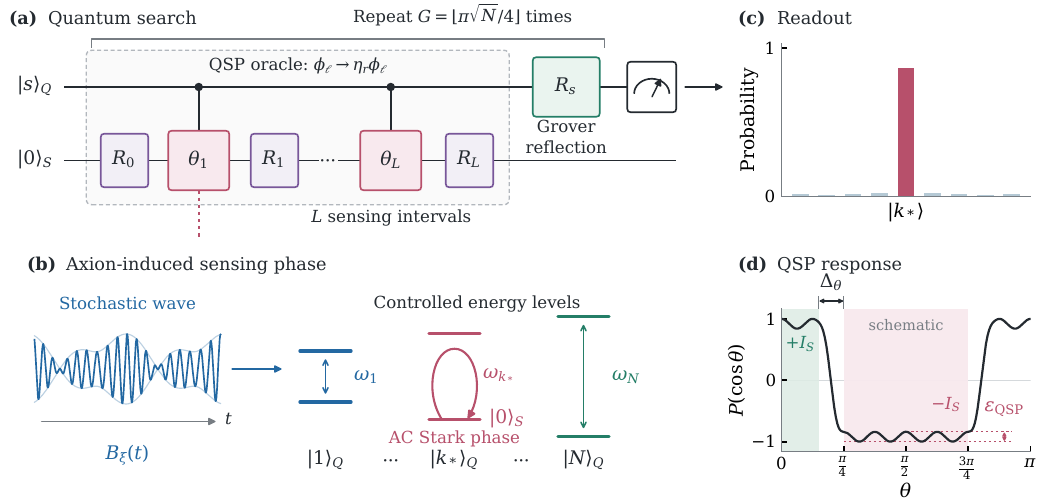}
 \caption{ \textbf{(a)} Protocol of Grover search for a stochastic DM signal. Each Grover iteration consists of a QSP oracle,
 containing $L$ ESUs interleaved with sensor rotations, 
 followed by a reflection on the register. 
 All QSP angles flip sign between oracle calls, with $\eta_r = (-1)^{r-1}$ on call $r$.
\textbf{(b)} Each register state $\ket{k}_Q$ changes the sensor's energy to $\omega_k$. 
The off-resonant stochastic field with effective coupling $B_\xi(t)$ drives the sensor to produce an AC Stark phase with negligible population transfer. 
The curved arrow illustrates the sensor
returning to $\ket{0}$ with an accumulated phase in the signal branch $k_\star$.
\textbf{(c)} Register measurement probabilities after $G$ Grover iterations, 
with the amplified signal bin $k_\star$ highlighted in red.
\textbf{(d)} The QSP response $P(\cos \theta)$ approximates $+1$ and $-1$ over the background and signal phase ranges, 
corresponding to oracle actions $+I_S$ and $-I_S$, respectively.
Here, $\Delta_\theta$ denotes the gap between background and signal phase ranges and $\epsilon_{\rm QSP}$ the maximum  approximation error.}
 \label{fig:sketch_protocol}
\end{figure*}

\section{AC Stark phase and stochastic fluctuations}
We first consider a signal in the monochromatic limit,
$B_\xi(t) = \sqrt{2} B_{\rm rms}\cos(m_a t+\varphi)$ with fixed $\varphi$. 
A general ESU is
\begin{equation}
 U_k^{(\xi)}
 = \zeta_k \ket0 \bra0 + \zeta_k^*\ket1 \bra1
   + \lambda_k\ket1 \bra0 - \lambda_k^*\ket0 \bra1,
 \label{eq:esu_components}
\end{equation}
with $\zeta_k= \sqrt{1-|\lambda_k|^2}\,e^{-i\theta_k}$ due to normalization.
For large detuning, $|\delta_k| \gg B_{\rm rms}$, 
and sufficiently slow switching of $\chi(t)$, transitions between the sensor levels are strongly suppressed,
$|\lambda_k|\ll1$.
The evolution is then nearly diagonal with
$U_k\simeq e^{-i\theta_kZ_S}$ and the two sensor levels acquire opposite
phases with small population transfer.
This is an adiabatic AC Stark phase. 
The smooth control function
$\chi(t-t_0)$ switches the effective DM sensor coupling on and off. 
When this switching is slow on the inverse detuning scale, 
a sensor initially in $\ket0$ follows the instantaneous eigenstate and returns to $\ket0$ when the coupling vanishes.
Its additional dynamical phase is then set by the time integral of the drive-induced energy shift. 
In the RWA, this eigenstate has energy
$(\delta_k/2) \sqrt{1 +2 B_{\rm rms}^2 \chi(t)^2/\delta_k^2}$.
Subtracting its free evolution term $\delta_k/2$ and integrating over the pulse, 
with $t_0=0$, gives
\begin{equation}
 \begin{aligned}
 \theta_k
 &\simeq \frac{\delta_k}{2} \int_0^T {\rm d} t \,
 \left[\sqrt{1+ \frac{2B_{\rm rms}^2 \chi(t)^2}{\delta_k^2}}-1\right]\\
 &\simeq \frac{B_{\rm rms}^2}{2\delta_k} \int_0^T {\rm d} t\, \chi(t)^2.
 \end{aligned}
 \label{eq:starkphase}
\end{equation}
The first line reproduces the result of Ref.~\cite{Allen:2025hdx}, and the second line is its large-detuning expansion. 
Note that $\theta_k$ is independent of $\varphi$ and proportional to $B_{\rm rms}^2$.
Therefore, a monochromatic drive produces the same $\theta_k$ in each ESU.

The stochastic DM wave is instead a linear combination of different momentum modes with random amplitudes and phases. 
Because the velocity dispersion is $v_0\sim10^{-3}$ in our halo, 
the kinetic energy $q_j/m_a = v_j^2/2$ give a frequency spread of order $10^{-6} m_a$.
Although narrow, 
this spectrum cannot be treated as monochromatic for $T \gg \tau_{\rm DM}$ where $\tau_{\rm DM} \sim (m_a v_0^2)^{-1}$ denotes the DM coherent time.
Interference between modes contains terms $\propto \cos[(q_j-q_l) t +\varphi_j - \varphi_l]$, causing the drive intensity at the sensor to vary continuously in time.
Here, the drive intensity is defined as the oscillation-period average of the squared amplitude of the DM signal, $\left\langle B_\xi(t)^2\right\rangle_{2\pi/m_a}$. 
Since the leading AC Stark shift is quadratic in the DM drive, 
this variation changes $\theta_k$ from ESU to ESU. 

The fluctuating DM signal can also lead to a nonzero population-transfer amplitude $\lambda_k$, which is small for modes far from resonance $m_av^2 \ll \delta_k$.
To suppress $\lambda_k$, 
we adopt the odd/even binning strategy of Ref.~\cite{Allen:2025hdx}.
The search band is divided into equal odd/even bins, and each bin consists of two intervals from $\omega_k \pm 3\beta/2$ to $\omega_k \pm \beta/2$ with width $\beta$ around the register-dependent frequency $\omega_k$, leaving a gap around resonance.
Choosing $\beta \gg m_a v_0^2$ keeps the entire DM speed profile away from $\omega_k$ (see Sec.~\ref{subsec:sm_modes}), 
suppressing the population transfer.

Long phase accumulation suppresses those fluctuations.
When $T \gg \tau_\text{DM}$, the ESU samples over many nearly uncorrelated DM signals so that the accumulated phase approach its ensemble average. 
With a smooth $\chi(t)$ and off-resonance interaction with every DM mode, 
the expected sensed phase is
\begin{equation}
 \overline \theta_k
 = \E[\theta_k^{(\xi)}]
 \simeq \frac{B_{\rm rms}^2}{2}
 \left[\int_0^T{\rm d} t \, \chi(t)^2 \right]
 \int  {\rm d} v \, \frac{f(v)}{\delta_k-m_a v^2/2}.
 \label{eq:meanstarkphase}
\end{equation}
The integral runs over the DM speeds, while $\delta_k-m_a v^2/2$ represents the velocity-dependent detuning between the sensor transition frequency and the driving frequency of the DM mode with speed $v$. 
Eq.~\eqref{eq:starkphase} is recovered when $f(v)$ peaks at a particular $v$, reproducing the monochromatic limit. 

Random initial phases cancel the interference terms in the ensemble mean, 
but a finite sensing interval leaves a residual phase variance,
$\sigma_{\theta,k}^2=\E[(\theta_k^{(\xi)}-\overline\theta_k)^2]$.
The relative
standard deviation satisfies
\begin{equation}
 \frac{\sigma_{\theta,k}}{|\overline\theta_k|}
 \simeq
 \sqrt{\frac{\tau_{{\rm eff},k}}{T}}.
 \label{eq:scaling}
\end{equation}
where $\tau_{{\rm eff},k} \simeq \tau_{\rm DM}$ is the effective correlation time determined by both the sensor response and the DM's lineshape (see SM Sec.~\ref{subsec:sm_long_time}). 
Intuitively speaking, while the expected phase accumulates coherently, the stochastic error adds incoherently such that both $\bar{\theta}_k$ and $\sigma_{\theta,k}^2$ are $\propto T$. 
Thus, increasing $T$ suppresses the relative error of $\theta_k$.

\section{Frequency selection with quantum signal processing}
Grover search usually requires a definite phase marking on the signal state $\ket{k_*}_Q$, 
but an ESU produces a phase that depends on the field strength and frequency. 
Since the signal and background ESU phases lie in separate ranges, 
QSP maps them to approximately $\pm \ket{k}_Q$, hence implementing the Grover oracle. 
The QSP is achieved by interleaving $L$ ESUs, $e^{-i\theta Z_S}$, with $L+1$ programmed sensor rotations, $e^{-i\phi_\ell X_S}$. 
For a monochromatic signal, every ESU contributes the same phase $\theta$. 
The QSP sequence can generate an at-most-degree-$L$ polynomial $P(\cos\theta)$ in the $\ket0$-to-$\ket0$ amplitude.
We program $X_S$ rotations with a particular list of $\phi_\ell$
so that $P(\cos\theta)$ is close to $-1$ for signal phases and $+1$ for
background phases, producing a unitary close to the desired $\pm I_S$ on the sensing qubit ~\cite{Allen:2025hdx,Martyn:2021eaf}.
Because the DM signal is stochastic, 
the actual oracle unitary is
\begin{align}
 Q_{k,r}^{(\xi)}& = e^{-i \eta_r \phi_L X_S} U_{k,r,L}^{(\xi)}\cdots
e^{-i \eta_r \phi_1 X_S} U_{k,r,1}^{(\xi)}
e^{-i \eta_r \phi_0 X_S},\nonumber\\
 \mathcal O_r^{(\xi)}&=\sum_k\ket{k}\bra{k}_Q\otimes Q_{k,r}^{(\xi)}.
 \label{eq:qsp}
\end{align}
The index $r$ labels oracle calls, 
with $\eta_r=(-1)^{r-1}$ specifying the sign applied to all QSP angles
and the last index of $U$ labels the ESUs within each oracle.
The QSP sequence length $L$ also depends on the chosen phase ranges.
A narrower gap between the signal and background ranges (see panel (d) of Fig.~\ref{fig:sketch_protocol}) requires a higher $L$ to approximate. 
Let the gap in $x=\cos\theta$ be $\Delta_x$.
A suitable polynomial can reach QSP approximation error $\epsilon_{\rm QSP}$ with $L=\mathcal{O}[\Delta_x^{-1}\log(1/\epsilon_{\rm QSP})]$
\cite{Allen:2025hdx}. 
Here, $\epsilon_{\rm QSP}$ is the largest
deviation of $P(x)$ from its target sign over the two ranges, with $P(x) =
\bra{0} Q_L (\theta) \ket{0}$ and $x = \cos \theta$, where $Q_L(\theta)$ denotes the QSP unitary while sensing a monochromatic signal generating an ESU phase of $\theta$. 

However, the polynomial response assumes that every ESU contributes the
same phase. 
For stochastic DM,  
the ESU sensing phase fluctuates, 
so the QSP response deviates from $P(\cos\theta)$ and acquires a stochastic error. 
Therefore, our numerical evaluation of Eq.~\eqref{eq:qsp} incorporates both the stochastic error and the QSP approximation error.
As a benchmark, we set $L=40$ and define the signal and background phase ranges as $\pi/4\leq|\theta|\leq3\pi/4$ and $|\theta|\leq3\pi/20$ respectively, so $\Delta_x \approx 0.184$ and $\epsilon_{\rm QSP} \approx 1.355\times10^{-3}$. 

If ESU phases fluctuate around the same mean with little correlation between intervals, 
their small perturbations add as a random walk. 
At fixed single-ESU variance $\sigma_\theta^2$ and $L \sigma_\theta^2 \ll 1$, 
the added stochastic error is therefore proportional to $L \sigma^2_\theta$ (see also Sec.~\ref{subsec:ESU-fluctuation}).
For $\bar\theta=\pi/4$, $m_a=10\,\mu\mathrm{eV}$, $T=140\,\mathrm{s}$, and $L = 40$, eq.~\eqref{eq:scaling} estimates an error of $L\sigma_{\theta,k}^2 \approx 5\times 10^{-5}$ while the numerical result is $6.31\times10^{-5}$. Therefore, the total ensemble-averaged error is $1.42\times10^{-3}$.

\begin{figure*}[!t]
 \centering
 \includegraphics[width=0.90\textwidth]{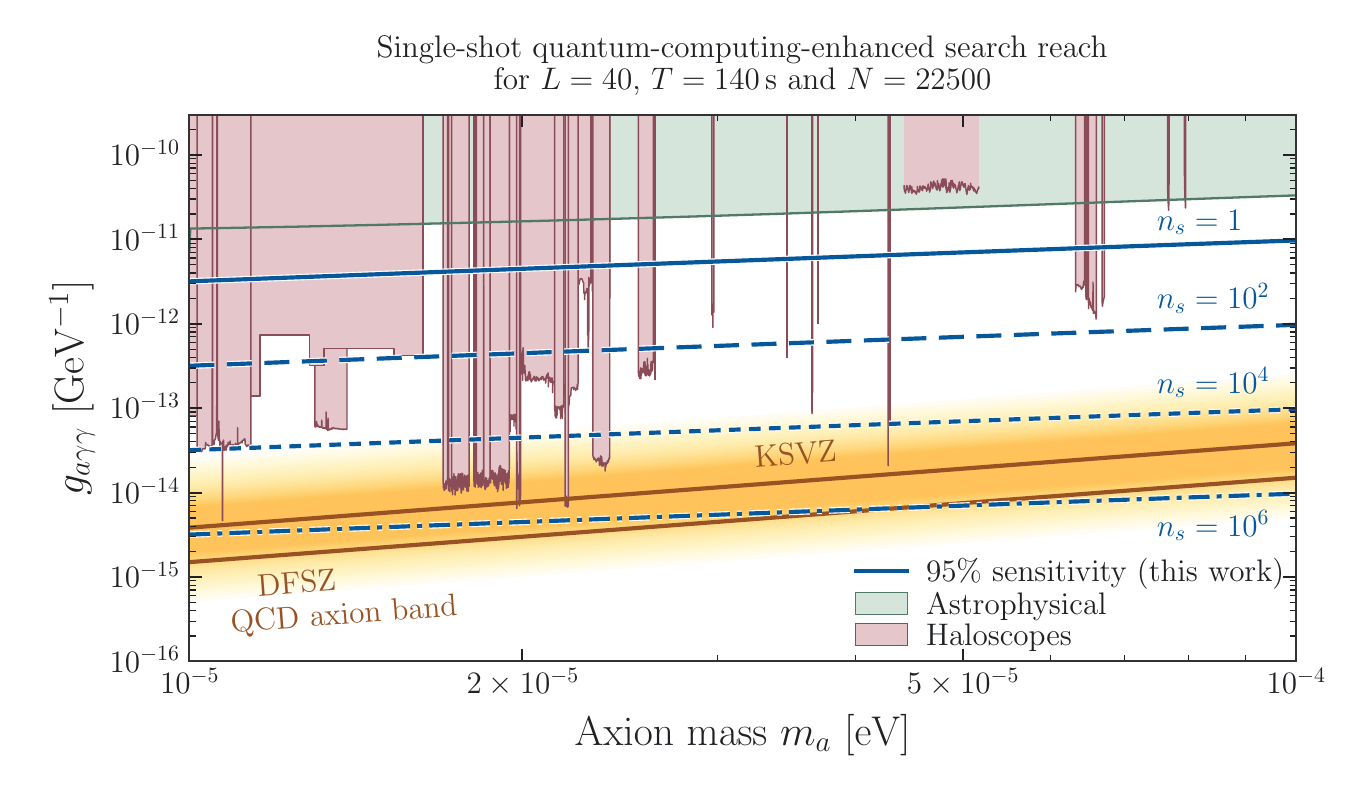}
 \caption{
 Projected single-search sensitivity for QSP sequence length $L=40$, number of frequency bins $N=22500$ (corresponding to $G=117$ Grover iterations), and fixed ESU duration $T=140\,$s.
 The blue curves (this work) show the projected 95\% sensitivity reach for the number of sensing qubits $n_s = 1$ (solid), $n_s = 10^2$ (long dashed), $n_s = 10^4$ (short dashed), and $10^6$ (dot-dashed).
 The gold band shows the conventional QCD axion model range, 
 with brown lines marking the KSVZ and DFSZ benchmarks.
 Green and red shaded regions show combined astrophysical and haloscope constraints~\cite{AxionLimits}, respectively.
 }
 \label{fig:axion}
\end{figure*}

\section{Grover search and detection probability}
We now apply the Grover search protocol to determine the probability of finding the signal frequency bin.
We initialize the register in a uniform superposition $\ket{s}_Q = N^{-1/2} \sum_{k=1}^N \ket{k}_Q$ and the sensor in $\ket0_S$. 
Each iteration applies the oracle in Eq.~\eqref{eq:qsp} and then a reflection $R_s=2 \ket{s} \bra{s}_Q-I_Q$ on the register, where $I_Q$ is the identity operator on the register. 
Hence, after $G$ Grover iterations, the register-sensor system is evolved by
\begin{equation}
 V_\xi(G) = \prod_{r=G}^{1}[(R_s\otimes I_S)
 \mathcal O_r^{(\xi)}].
 \label{eq:grover}
\end{equation}
For an ideal oracle ($-I_S$ on the signal bin and $+I_S$ elsewhere), after $G = \lfloor\pi\sqrt{N}/4\rfloor$ iterations, its signal-bin readout probability is nearly maximized.

However, the oracle is always imperfect because of both the QSP approximation error and the DM-induced stochastic error. 
Repeating the same QSP sequence can accumulate coherent errors and reduce the search success probability. 
To suppress the QSP approximation error, 
we alternate between the original QSP sequence and a sequence with every QSP angle's sign flipped.
In the monochromatic limit, this sign flip implements the adjoint QSP unitary, 
so the dominant QSP approximation error cancels in each pair of Grover iterations.
Only a subleading error remains as
\begin{equation}
\epsilon_{\rm pair}
\sim
\frac{\epsilon_{\rm QSP}}{N}
\end{equation}
in each alternating-sign pair.
Therefore, over $\sqrt{N}$ iterations, the accumulated QSP approximation error improves from $\mathcal{O}(N\epsilon_{\rm QSP})$ to $\mathcal{O}(\epsilon_{\rm QSP})$ with the our new protocol.

However, stochastic errors are independent between oracle calls and cannot be canceled. 
To account for their effects on both the register phases and the associated sensor states, 
our simulation evolves the register and sensor jointly through $G$ iterations.
The derivation of this cancellation and a comparison of the search success probability with and without the alternation, including phase fluctuations, 
are given in Sec.~\ref{subsec:Alternating-QSP}.

For a given DM configuration $\xi$, 
the final density operator is $\rho_\xi(G) = V_\xi(G) \ket{s,0} \bra{s,0} V_\xi(G)^\dagger$, 
where $\ket{s,0} = \ket{s}_Q\otimes \ket0_S$. 
Reading out the register state and summing
over both sensor outcomes gives the probability of the target label $k_*$:
\begin{equation}
 P_{\rm bin}(\xi) = 
 \operatorname{Tr}[(\ket{k_*}\bra{k_*}_Q\otimes I_S)
    \rho_\xi(G)].
 \label{eq:Pbinhistory}
\end{equation}
Since $\xi$ is totally random, averaging the conditional probability in Eq.~\eqref{eq:Pbinhistory} over its distribution $p_{\rm halo}(\xi) {\rm~d} \xi = \prod_j
\left[A_j e^{-A_j^2 / 2} {\rm ~d} A_j {\rm ~d} \varphi_j/2 \pi\right]$ gives the expected success probability of one search
\begin{equation}
 \overline P_{\rm bin}
 = \int{\rm d}\xi\,p_{\rm halo}(\xi)P_{\rm bin}(\xi).
 \label{eq:PbinAver}
\end{equation}
In this work, we use Monte Carlo sampling over the initial DM configurations $\xi$ to numerically evaluate the expected success probability $\overline P_{\rm bin}$. 
We interpret $\overline P_{\rm bin}$ as the probability that a single search correctly identifies the signal bin assuming the Standard Halo Model (see Sec.~\ref{subsec:sm-ensemble}).

To determine the weakest detectable signal at each DM mass, 
we fix $T$ and the circuit parameters and decrease $B_{\rm rms}$. 
Since $|\overline{\theta}_k| \propto B_{\rm rms}^2$ in Eq.~\eqref{eq:meanstarkphase}, 
the expected sensing phase eventually falls below the lower edge of the QSP signal phase range, 
causing the signal-bin oracle unitary to depart rapidly from $-I_S$ and reducing the search success probability.
The projected 95\% sensitivity can then be defined by the first crossing of 
\begin{equation}
\overline P_{\rm bin}(B_{\rm rms})=0.95
\label{eq:Sensitivitycriterion}
\end{equation}
as $B_{\rm rms}$ decrease.
In other words, the crossing gives the minimum detectable $B_{\rm rms}$ with a 95\% success probability.

\section{Axion sensitivity and sensing time}
To demonstrate the protocol, we consider axion masses $m_a \in [m_{a, \rm min}, m_{a, \rm max}] = [10^{-5}\,\mathrm{eV}, 10^{-4}\,\mathrm{eV}]$. 
This mass window is of interest for QCD axion dark matter and is targeted by the transmon-qubit search proposed in Ref.~\cite{Chen:2024aya}.
For the axion–photon interaction $\mathcal{L}_{a \gamma} = -g_{a \gamma \gamma} a \tilde{F}_{\mu \nu} F^{\mu \nu} / 4$, 
an applied magnetic field $B_0$ induces an oscillating electric field proportional to the axion field. 
The resulting transverse drive of the transmon takes the form of Eq.~\eqref{eq:DMsensorresponse}, 
with $B_{\rm rms}$ determined by the axion–photon coupling $g_{a\gamma\gamma}$ and the device parameters. 
In the harmonic-transmon approximation, this yields~\cite{Chen:2024aya}
\begin{equation}
 B_{\rm rms}
 =
 \frac{|g_{a\gamma\gamma}|B_0}{m_a}
 |\kappa\cos\Theta|\,d\sqrt{\frac{C \omega_k \rho_{\rm DM}}{2}}.
 \label{eq:transmonrms}
\end{equation}
Here, $B_0$ is the applied static magnetic field, $C$ the capacitance, $\omega_k$ the register-dependent sensing frequency 
and $d$ the effective electrode separation. 
The dimensionless factor
$\kappa$ describes the sensor's electromagnetic response function, and $\Theta$ is the angle between the induced electric field and the sensitive axis. For simplicity, we will assume that $\kappa = 1$ and $\cos \Theta = 1$.
For the benchmark we take $B_0=5\,\mathrm T$,
$C=0.1\,\mathrm{pF}$, $d=100\,\mu\mathrm m$, with local axion density
$\rho_{\rm DM}=0.4\,\mathrm{GeV\,cm^{-3}}$.

We choose the frequency bin width $\beta$ conservatively to be one-order-of-magnitude larger than the DM bandwidth at $m_{a,\rm max}$, i.e., $\beta=10^{-9}\,\mathrm{eV}$. 
This choice ensures that the DM frequency spectrum remains off resonance in the appropriate run.
With the odd/even binning strategy, 
each run takes half of the $9\times10^4$ basic intervals, two per label $k$, giving a total number of $N=(m_{a, \rm max} - m_{a, \rm min})/(4\beta)=22500$ labels and requiring $G=117$ Grover iterations. 
We use an $L=40$ QSP sequence and fix $T=5 \times 10^5 \tauDM(m_{a,{\rm min}})=140\,\mathrm{s}$ so that the total search time is $\sim 2 G L T \approx 364\,$h. 
We choose $L = 40$ as, for our $N$, it is roughly the smallest $L$ for which $\overline P_{\rm bin} = 95\%$ is attainable, with the sensitivity mostly determined by $\overline\theta_k$ (see Secs.~\ref{sm:qsp} and \ref{subsec:sm-ensemble}).
We also discuss how a repeated search trades repeated sampling for a shorter sensing time, reducing the overall timing complexity, in Sec.~\ref{subsec:sm-repetitions}. 

At each mass, we numerically evolve the density operator of the control qubit and sensor and calculate the expected success probability in Eq.~\eqref{eq:PbinAver}. 
The halo distribution and evolution procedure are specified in SM Secs.~\ref{sm:field} and \ref{sm:inference}. 
Applying the criterion in Eq.~\eqref{eq:Sensitivitycriterion} then yields $B_{{\rm rms},95}$, 
the drive strength corresponding to the projected 95\% sensitivity. 
With these choices, Fig.~\ref{fig:axion} gives $|g_{a\gamma\gamma,95}| \simeq 3.18 \times10^{-12}$-- $9.71 \times10^{-12}\,\mathrm{GeV^{-1}}$ over the mass range $m_a \in [m_{a, \rm min}, m_{a, \rm max}]$.
The projected sensitivity substantially improves on existing astrophysical bounds, 
while the broadband search covers extensive mass regions unexplored by conventional narrowband cavity experiments.
The dashed curves illustrate the further sensitivity improvement possible with entangled sensing qubits. 
For $n_s$ entangled sensors, the phases add to $n_s\theta_k$. 
Since $\theta_k\propto g_{a\gamma\gamma}^2$, 
ideal coherent accumulation reduces the coupling threshold by a factor of $1/\sqrt{n_s}$. 
Although $n_s = 10^{6}$ seems aggressive, our benchmark is realistic given the long coherence and control sufficient to support the long sensing time $T=140\,{\rm s}$ on the register.
Preferably, the register qubits are already logical and error-corrected by many physical qubits.
Long-term efforts toward million-qubit processors~\cite{GoogleQuantumAI:Roadmap} also motivate considering $10^6$ entangled sensors.
Provided that collective coherence and control can be maintained throughout the search,
this scaling would enable the search to probe the KSVZ QCD axion models, as well as much of the DFSZ band, in this mass range.

\section{Conclusion}
In this work, we demonstrate that Grover search can provide a sensitive broadband probe of wavelike DM, 
accounting for both the DM-induced stochastic fluctuations and the imperfect QSP. 
At a fixed mean sensing phase, longer sensing durations suppress the stochastic error.
Alternating the signs of all QSP control angles further reduces the coherent accumulation of the QSP approximation error. 
Together, these measures substantially improve the search success probability and enable reliable identification of the DM signal.

Applying this protocol to axion DM, 
our benchmark with ESU time $T=140\,\mathrm{s}$ achieves a projected 95\% sensitivity beyond existing astrophysical bounds. 
The broadband search probes unexplored parameter space across the $10$–$100\,\mu\mathrm{eV}$ mass range. 
The sensitivity can be further extended with entangled sensing qubits.
With $10^6$ entangled sensing qubits, one would further extend this sensitivity to standard QCD axion benchmarks.

\bigskip

\let\section\oldsection

\acknowledgments 
We thank Soonwon Choi for helpful discussions. In finalizing the draft, we became aware of an independent study on a similar topic. We would like to acknowledge Roni Harnik and Zachary Bogorad for coordinating the concurrent arXiv submission of both manuscripts. Y.B. and L-T. W. acknowledge the support from the Department of Energy grant DE-SC0009924, and the 2026 Joint Task Force Initiative (JTFI) Partnerships for Emerging Technologies seed fund. Y.C.'s work is supported by the National Research Foundation of Korea (NRF) Grant RS-2023-00211732, 
by the Samsung Science and Technology Foundation under Project Number SSTF-BA2302-05, by the POSCO Science Fellowship of POSCO TJ Park Foundation, and by the NRF Grant RS-2024-00405629. L.J.'s work is supported by the ARO(W911NF-23-1-0077), ARO MURI (W911NF-21-1-0325), AFOSR MURI (FA9550-21-1-0209, FA9550-23-1-0338), ONR MURI (N000142612102), NSF (OSI-2326767, OSI-2426975, EEC-2550064, OSI-2553578, OSI-2553619).

\bibliographystyle{apsrev4-1}
\bibliography{references}

\clearpage
\onecolumngrid
\raggedbottom
\setlength{\tabcolsep}{6pt}
\setcounter{equation}{0}
\setcounter{figure}{0}
\setcounter{table}{0}
\setcounter{section}{0}
\setcounter{subsection}{0}
\setcounter{secnumdepth}{2}
\setcounter{page}{1}
\counterwithout{equation}{section}
\renewcommand{\thesection}{S\arabic{section}}
\renewcommand{\thesubsection}{\thesection.\Alph{subsection}}
\makeatletter
\def\p@subsection{}
\makeatother
\renewcommand{\theequation}{S\arabic{equation}}
\renewcommand{\thefigure}{S\arabic{figure}}
\renewcommand{\thetable}{S\arabic{table}}
\renewcommand{\thepage}{S\arabic{page}}

\begin{center}
\textbf{\large Quantum Computing Enhanced Search for Wavelike Dark Matter}\par
\vspace{0.05in}
{\itshape\large Supplemental Material}\par
\vspace{0.05in}
Yunjia Bao, Yu Cheng, Liang Jiang, Lian-Tao Wang
{}
{}
\end{center}

This Supplemental Material demonstrates how the error from the DM signal's stochastic phase fluctuations propagates into the frequency search.  
Sec.~\ref{sm:field} specifies the field and the average over its unknown amplitudes and phases. 
Sec.~\ref{sm:esu} derives the response during one elementary sensing unit (ESU), 
the conditions for small transitions, 
and the phase statistics that quantify intensity averaging.
Sec.~\ref{sm:qsp} then explains how quantum signal processing (QSP) uses the phase to distinguish one register label from the others,
and how the stochastic errors affect the search. The expected success probability of finding the target label at $95\%$ gives the search sensitivity as discussed in Sec.~\ref{sm:inference}.

\section{Stochastic DM field and coherence time}\label{sm:field}

The stochastic axion field $a_\xi(t)$ is a linear combination of velocity modes, 
as expressed in Eq.~\eqref{eq:stochasticfield}. 
For DM mass $m_a$, a mode of speed $v_j$ oscillates at $m_a+q_j$, 
where $q_j=m_av_j^2/2$.
Each mode has an amplitude proportional to $A_j\sqrt{f(v_j)\Delta v}$, 
where $A_j$ is Rayleigh distributed following a probability distribution $p(A_j) = A_j \exp(-A_j^2/2)$, and $\varphi_j$ is uniformly distributed on $[0,2\pi)$~\cite{Foster:2017hbq}. 
These coefficients are mutually independent, 
and each configuration $\xi=\{A_j,\varphi_j\}$ yields one realization of the DM waveform.
The ensemble average $\E$ of a function $g(\xi)$ is $\int \mathrm{d} \xi\, p_{\rm halo}(\xi) g(\xi)$ with the joint probability measure
\begin{equation}
    \mathrm{d} \xi \, p_{\rm halo }(\xi)
    =
    \prod_j \left[A_j e^{-A_j^2 / 2} 
    \mathrm{~d} A_j 
    \frac{\mathrm{~d} \varphi_j}{2 \pi}\right].
\end{equation}

The root-mean-square (RMS) amplitude of the DM waveform is obtained by squaring the mode expansion and averaging over the random amplitudes and phases.
The resulting expression contains diagonal terms, with the same mode appearing twice, 
and cross terms involving two distinct modes. 
Because the phases of different modes are independent and uniformly distributed, 
all cross terms vanish upon averaging. For each diagonal term, 
we obtain $\E[A_j^2]=2$ and $\E[\cos^2\varphi_j] = 1/2$. 
Since the DM speed distribution satisfies $\int_0^{\infty} f(v) \mathrm{d} v=1$, we obtain
\begin{equation}
 B_\xi(t)=\frac{m_aB_{\rm rms}}{\sqrt{\rho_{\rm DM}}}a_\xi(t),\qquad
 \E [a_\xi(t)^2]=\frac{\rho_{\rm DM}}{m_a^2},\qquad
 \E [B_\xi(t)^2]=B_{\rm rms}^2.
 \label{eq:sm_fieldrms}
\end{equation}
Here, $\rho_{\rm DM}$ is the local DM energy density, and $B_\xi(t)$ is the effective DM sensor coupling in Eq.~\eqref{eq:DMsensorresponse},
whose RMS coupling strength is $B_{\rm rms}$.

We adopt the Standard Halo Model, 
in which the DM velocities follow an isotropic Maxwell–Boltzmann distribution in the Galactic frame, truncated at the Galactic escape speed~\cite{Freese:2012xd,Baxter:2021pqo}. 
The normalized three-dimensional velocity distribution is
\begin{equation}
 F_{\rm gal} (\mathbf u)
 =
 \frac{e^{-u^2/v_0^2}}
 {\mathcal N_{\rm esc}\pi^{3/2}v_0^3}
 \Theta (v_{\rm esc}-u),
 \qquad 
 \mathcal N_{\rm esc}
 =
 \operatorname{erf}\!\left(\frac{v_{\rm esc}}{v_0}\right)
 -
 \frac{2v_{\rm esc}}{\sqrt\pi v_0}
  e^{-v_{\rm esc}^2/v_0^2},
 \label{eq:sm_galactichalo}
\end{equation}
where $u=|\mathbf u|$ is the DM speed in the Galactic frame, 
$v_0$ sets the width of its distribution, 
and $v_{\rm esc}$ is the Galactic escape speed at which the Heaviside function $\Theta$ truncates the velocity distribution. 
The factor $\mathcal N_{\rm esc}$ keeps $F_{\rm gal}(\mathbf u)$ normalized once the high speed tail is removed.

The sensor moves through this distribution with a laboratory velocity
$\mathbf v_{\rm obs}$.
A velocity $\mathbf v$ relative to the laboratory
therefore corresponds to $\mathbf v+\mathbf v_{\rm obs}$ in the Galactic frame, 
so $F_{\rm lab}(\mathbf v)=F_{\rm gal}(\mathbf v+\mathbf v_{\rm obs})$.
Because the mode frequency only depends on its speed after the DM mass is fixed, 
the relevant mode spectrum is determined by the DM's speed distribution
which can be obtained by angular integration, $f(v)=v^2\int{\rm d}\Omega\,
F_{\rm lab}(\mathbf v)$. This yields
\begin{equation}
 f(v)
 =
 \frac{2\pi v^2}{\mathcal N_{\rm esc}\pi^{3/2}v_0^3}
 \int_{-1}^{\mu_{\max}} \mathrm{d} \mu\, 
 e^{-(v^2+v_{\rm obs}^2+2vv_{\rm obs}\mu)/v_0^2},
 \qquad
 \mu_{\rm max}
 =
 \min\!\left(1,
 \frac{v_{\rm esc}^2 - v^2 - v_{\rm obs}^2}{2 v v_{\rm obs}}\right),
\end{equation}
where $\mu$ is the cosine of the angle between $\mathbf v$ and $\mathbf v_{\rm obs}$.
Evaluating this integral gives
\begin{equation}
 f(v) = 
 \frac{v}{\mathcal N_{\rm esc}\sqrt\pi v_0 v_{\rm obs}}
 \begin{cases}
 e^{-(v-v_{\rm obs})^2/v_0^2}-e^{-(v+v_{\rm obs})^2/v_0^2},
 &0\leq v\leq v_{\rm esc}-v_{\rm obs},\\[2pt]
 e^{-(v-v_{\rm obs})^2/v_0^2}-e^{-v_{\rm esc}^2/v_0^2},
 &v_{\rm esc}-v_{\rm obs}<v\leq v_{\rm esc}+v_{\rm obs},\\[2pt]
 0,&\text{otherwise}.
 \end{cases}
 \label{eq:sm_labspeed}
\end{equation}
At low laboratory speeds, every direction is allowed. Between $v_{\rm esc}-v_{\rm obs}$ and $v_{\rm esc}+v_{\rm obs}$, 
some directions
would correspond to a Galactic speed above the cutoff and are excluded.
Above the latter speed no modes remain.

We adopt $v_0=220$, $v_{\rm obs}=232$, and $v_{\rm esc}=544$ in
$\mathrm{km\,s^{-1}}$. 
In the lab frame, the maximum DM speed is
$v_{\max}=v_{\rm esc}+v_{\rm obs}$, which sets the largest kinetic frequency shift, $q_{\max}=m_av_{\max}^2/2$.
For $m_a=100\,\mu\mathrm{eV}$, this gives $q_{\max}\approx3.35\times10^{-10}\,\mathrm{eV}$.
The occupied DM spectrum therefore extends from $m_a$ to $m_a+q_{\max}$.

Following Ref.~\cite{Cheong:2024ose}, 
we take the DM coherence time in terms of the halo speed distribution $f(v)$ defined in Eq.~\eqref{eq:sm_labspeed} as
\begin{equation}
 \tau_{\mathrm{DM}} \simeq 
 \frac{2 \pi}{m_a} \int \frac{\mathrm{d} v}{v} f(v)^2 .   
\end{equation}
Evaluating this integral gives
\begin{equation}
 \tauDM(m_a) \approx 0.389
 \frac{2 \pi}{m_av_0v_{\rm obs}}
 \label{eq:smtauDM}
\end{equation}
For DM mass $m_a = 10^{-5}\,$eV, the coherence time is around $\tau_{\rm DM}(10^{-5}{\rm eV}) \approx 2.83 \times 10^{-4}\,$s.

\section{Sensor evolution and phase fluctuations}
\label{sm:esu}

The stochastic DM field induces both phase accumulation and population transfer in the sensing qubit. 
We first identify when an ESU can be approximated as a pure phase rotation and then calculate the phase statistics relevant to the search and their dependence on the sensing time.

\subsection{Controlled Hamiltonian and ESU}

The bare sensor has Hamiltonian $H_0=\omega_S Z_S/2$, where $\omega_S$ is its physical level separation.
An index register label $\ket k_Q$ sets the splitting to $\omega_k$.
During a sensing window that begins at $t_0$ and lasts for a duration $T$, 
the joint lab-frame Hamiltonian is given by Eq.~\eqref{eq:sensor}.
This controlled Hamiltonian can be realized, for example, 
by introducing an Ising interaction $\sum_i J_i Z_{Q,i}Z_S/2$, 
which gives $\omega_k=\omega_S+\sum_i(-1)^{k_i}J_i$, 
where $k_i$ denotes the bit value of the $i$th register qubit and $J_i$ is its coupling to the sensor~\cite{Allen:2025hdx}.

To separate the oscillating DM signal from the sensor frequency, 
define $R_a (t)=e^{i m_a t Z_S/2}$ and transform the state as
$\ket{\psi_{\rm rot}(t)} = R_a(t)\ket{\psi(t)}$.
The corresponding  Hamiltonian then transforms as
\begin{align}
 H_{{\rm rot},k}^{(\xi)}(t)
 &= R_a(t) H_k^{(\xi)}(t) R_a^\dagger(t)
   +i \dot R_a(t) R_a^\dagger(t) \nonumber\\
 &= \frac{\delta_k}{2}Z_S+
 \frac{B_{\rm rms} \chi(t-t_0)}{2}
 \sum_j A_j \sqrt{f(v_j)\Delta v}\,
 \left\{\left[e^{-i(q_jt+\varphi_j)}
 +e^{i[(2m_a+q_j)t+\varphi_j]}\right]\sigma_+
 +\mathrm{H.c.}\right\},
\end{align}
where $\delta_k = \omega_k - m_a$ and $\sigma_\pm$ denotes the ladder operators.
Each mode contains a slowly rotating term with frequency $q_j=m_a v_j^2/2$ and a rapidly rotating term near $2m_a$. 
The latter is negligible as it averages out over the sensing time, 
and the rotating wave approximation (RWA) therefore yields $\widetilde H^{(\xi)}_k(t)$ in Eq.~\eqref{eq:rotatingH}.

Removing the free precession generated by $\delta_kZ_S/2$ gives the ESU unitary
\begin{equation}
 U_k^{(\xi)}(t_0,T)\simeq
 e^{i\delta_k(t_0+T)Z_S/2}
 \mathcal T\exp\!
 \left[-i\int_{t_0}^{t_0+T}
 \widetilde H_k^{(\xi)}(t){\rm d} t\right]
 e^{-i\delta_kt_0Z_S/2}.
 \label{eq:smrwaesu}
\end{equation}
Equivalently, we can remove the free precession phase by transforming to the interaction picture. 
Applying the unitary transformation $e^{i\delta_k t Z_S/2}$ eliminates the free term from $\widetilde H_k^{(\xi)}$ and yields the interaction Hamiltonian
\begin{equation}
 H_{I,k}^{(\xi)}(t)=\frac{B_{\rm rms}\chi(t-t_0)}2
 \sum_j A_j\sqrt{f(v_j)\Delta v}
 \left[e^{i(\Delta_{k,j}t-\varphi_j)}\sigma_+
       +e^{-i(\Delta_{k,j}t-\varphi_j)}\sigma_-\right].
 \label{eq:sminteraction}
\end{equation}
Here, $\Delta_{k,j}=\delta_k-q_j$ is the detuning between the sensor and the $j$th mode. 
The ESU evolution operator can then be obtained directly from this interaction Hamiltonian as
$U_k^{(\xi)}(t_0,T) = 
\mathcal T\exp[-i\int_{t_0}^{t_0+T}
H_{I,k}^{(\xi)}(t){\rm d} t]$.

For a specified DM waveform, 
this propagator can be evaluated through direct numerical integration of the two-level Schrödinger equation, 
yielding the full ESU matrix. See Sec.~\ref{subsec:sm-direct} for more details.
Combining these matrices with the programmed QSP sequence and register reflections then gives the complete Grover iteration.

To study the ESU of a stochastic signal with good analytical control, 
we now expand in the weak-signal limit.
For an initial $\ket0$, 
we define the return amplitude $\zeta_k=\bra0U_k^{(\xi)} \ket0$ and transition amplitude $\lambda_k = \bra1U_k^{(\xi)}\ket0$. 
The traceless two-level interaction Hamiltonian generates an $SU(2)$ propagator, 
\begin{equation}
 U_k^{(\xi)}
 =
 \begin{pmatrix}
 \zeta_k&-\lambda_k^*\\
 \lambda_k& \zeta_k^*
 \end{pmatrix}.
\end{equation}
with $|\zeta_k|^2+|\lambda_k|^2=1$.
To second order in the signal, the Dyson expansion gives
\begin{equation}
 U_k^{(\xi)}\ket0 = \ket0
 -i\int_{t_0}^{t_0+T}{\rm d} t\,H_{I,k}^{(\xi)}(t)\ket0
 -\int_{t_0}^{t_0+T}{\rm d} t_1\int_{t_0}^{t_1}{\rm d} t_2\,
 H_{I,k}^{(\xi)}(t_1)H_{I,k}^{(\xi)}(t_2)\ket0 
 +\cdots.
 \label{eq:smdyson}
\end{equation}
Because $H_{I,k}^{(\xi)}$ is always off-diagonal for the sensor, $\lambda_k$ is linear in the drive strength, 
while the perturbation to $\zeta_k$ is quadratic in the signal. 
Their dependence on the sensing duration and detuning is described by
\begin{equation}
 \mathcal F_T(\Delta)=\int_0^T\mathrm{d} t\, \chi(t)e^{i\Delta t},
 \qquad
 \mathcal K_{jl}=\int_0^T{\rm d} t_1\int_0^{t_1}{\rm d} t_2\,
 \chi(t_1)\chi(t_2)e^{i\Delta_{k,j}t_1-i\Delta_{k,l}t_2}.
 \label{eq:smfilters}
\end{equation}
The time here is measured from the start of the sensing process.
The function $\mathcal F_T$ describes a single transition, while $\mathcal K_{jl}$ describes a return through mode $l$ at the earlier time $t_2$ and mode $j$ at the later time $t_1$. 
Substitution into the Dyson expansion gives
\begin{align}
 \lambda_k^{(1)}
 ={}&-\frac{iB_{\rm rms}}2
 \sum_j A_j\sqrt{f(v_j)\Delta v}\,
 e^{i\varphi_j-i\Delta_{k,j}t_0}\mathcal F_T(\Delta_{k,j})^*,
 \label{eq:sm_transition_kernel}\\
 \theta_k^{(2)}
 ={}&\frac{B_{\rm rms}^2}{4}\operatorname{Im}
 \sum_{j,l}A_jA_l\sqrt{f(v_j)f(v_l)}\,\Delta v\,
 e^{i[\varphi_l-\varphi_j+(\Delta_{k,j}-\Delta_{k,l})t_0]}
 \mathcal K_{jl},
 \label{eq:sm_phase_kernel}
\end{align}
where we assume $\lambda_k$ is small.
Over a sufficiently long sensing time $T$, 
the leading phase can accumulate to order unity while the population transfers remain suppressed, as we justify below. 
In this regime, the ESU matrix is approximately diagonal, 
and can be written as
$U_k\simeq e^{-i\theta_k^{(2)}Z_S}$, 
which describes the accumulated phase rotation.

As shown in Eq.~\eqref{eq:sm_transition_kernel}, the sensor transition amplitude is proportional to $\mathcal F_T(\Delta)$. Following Ref.~\cite{Allen:2025hdx}, we suppress this transition error by smoothly turning the DM–sensor coupling on and off using an envelope function $\chi(t)$. Writing $\chi(t)=w(t/T)$ and defining $x=t/T$, we assume that $w$ and its derivatives up to order $p-1$ vanish at $x=0$ and $x=1$. For $\Delta\neq0$, integrating Eq.~\eqref{eq:smfilters} by parts $p$ times then gives
\begin{equation}
 |\mathcal F_T(\Delta)|
 \leq\frac{T}{(|\Delta|T)^p}
       \int_0^1|w^{(p)}(x)|{\rm d} x,
 \qquad p=1,2,\ldots.
 \label{eq:sm_switching_bound}
\end{equation}
When $|\Delta|T \gg 1$, the oscillatory contributions to $\mathcal{F}_T(\Delta)$ largely cancel. 
Smoothly turning the coupling on and off eliminates the boundary terms that would otherwise weaken this cancellation. 
We adopt the same envelope function $\chi(t)=w(t/T)$ as in Ref.~\cite{Allen:2025hdx}, 
normalized to unit peak as
\begin{equation}
 w(x)
 =
 \begin{cases}
 \exp\left[4-\dfrac{1}{x(1-x)}\right],&0<x<1,\\0,&\text{otherwise}.
 \end{cases}
 \label{eq:sm_pulse}
\end{equation}
All derivatives of $w$ vanish at both endpoints, so the bound in Eq.~\eqref{eq:sm_switching_bound} applies at every order $p$. 
The function $w(t/T)$ also affects $\theta_k$ and its variance through the dimensionless integrals
\begin{equation}
 \eta_p=\int_0^1 w(x)^p{\rm d} x.
 \label{eq:sm_pulse_weights}
\end{equation}
The long-duration mean value of $\theta_k$ depends on $\eta_2=0.2891131037$, 
whereas the variance involves $\eta_4=0.2122085942$, as derived in
Sec.~\ref{subsec:sm_long_time}.

\subsection{Odd/even frequency binning}
\label{subsec:sm_modes}

To suppress transition errors arising from the off-diagonal elements of the ESU matrix, 
the frequency bins must be sufficiently separated from the DM spectrum. 
Because $m_a$ is unknown, 
the control frequencies and corresponding acceptance bin width must be arranged to maintain this separation across the entire search band. 
We therefore adopt the odd/even binning scheme of Ref.~\cite{Allen:2025hdx}, 
and choose the bin width to accommodate the finite bandwidth of the DM halo.

\begin{figure}[!t]
 \centering
 \includegraphics[width=0.95\linewidth]{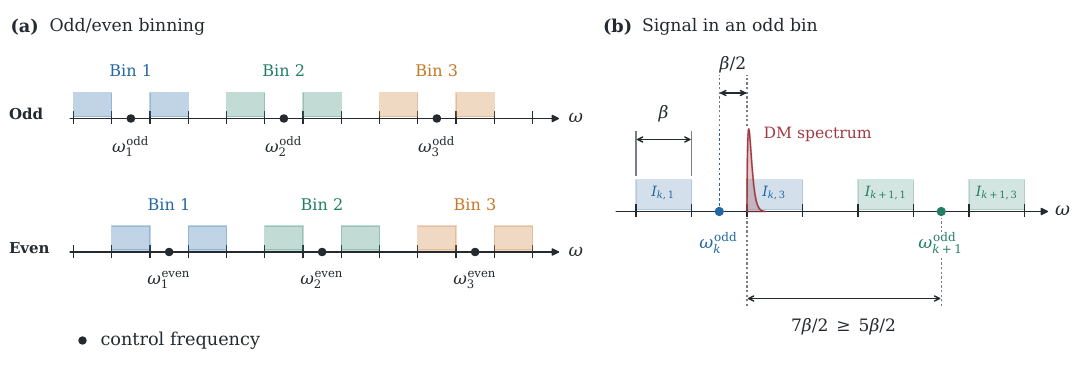}
 \caption{Odd/even binning and an example signal placement.
 \textbf{(a)} Schematic of the construction in Fig.~3(d) of
 Ref.~\cite{Allen:2025hdx}, redrawn here. Intervals of the same color
 on each row form one bin and share a control frequency (dot) in their
 gap. For DM, its detected frequency is typically $\omega \approx m_a$.
 Odd and even runs are searched separately and together cover
 the band. 
 \textbf{(b)} Two neighboring odd bins, each comprising two intervals
 of width $\beta$. A DM signal with a typical frequency $m_a$ at the left endpoint of $I_{k,3}$ has a detuning of
 $m_a-\omega_k^{\rm odd}=\beta/2$. The next-to-nearby control frequency in the same run
 is $7\beta/2$ away, consistent with the minimum $5\beta/2$ detuning required by QSP.
 The red curve shows the halo frequency density, with arbitrary height
 and horizontal width drawn to scale, taking $\beta=10^{-5}m_a$
 for this illustration such that
 $q_{\max}\simeq0.335\beta$ for the halo parameters in
 Sec.~\ref{sm:field}.}
 \label{fig:sm-geometry}
\end{figure}

The search band is divided into equal intervals of width $\beta$ and searched in two separate, interleaved runs. 
In each run as shown in Fig.~\ref{fig:sm-geometry}, a register label $k$ represents two nonadjacent bins, 
while the corresponding control frequency $\omega_k$ lies in the gap between them. 
The binning used in the second run is shifted by one interval relative to that used in the first, thereby covering the bins omitted from the first run.
For a search band starting at $\omega_{\min}$ with a total width
$4N\beta$, define groups of four intervals of width $\beta$:
\begin{equation}
 I_{k,a}
 =
[\omega_{\min}+(4(k-1)+a-1)\beta,
\omega_{\min}+(4(k-1)+a)\beta],
 \quad k=1,\ldots,N,\quad a=1,2,3,4.
 \label{eq:sm_primitive_bins}
\end{equation}
Here, $k$ labels the group, and $a$ labels a bin within that group, with the final endpoint included in the full search band. 
In the odd run, label $k$ covers $I_{k,1}$ and $I_{k,3}$, 
with their common control frequency at $\omega_{\min}+(4(k-1)+3/2)\beta$. 
The even run uses $I_{k,2}$ and
$I_{k,4}$, with control frequency at
$\omega_{\min}+(4(k-1)+5/2)\beta$~\cite{Allen:2025hdx}.
Together, the two runs cover the entire search band. 
A signal lying close to the control frequency in one run, which may violate the assumed $B \ll \delta_k$, is therefore assigned to an off-resonant bin in the other run such that a successful search is guaranteed.

Neglecting the DM's bandwidth,
the DM signal's detuning from the nearby assigned control frequency lies approximately between $\beta/2$ and $3\beta/2$,
while that from every other bin is at least $5\beta/2$.
With a weak DM signal,
having a minimum detuning of $\beta/2$ suppresses transitions between levels.
Then, the gap between $3\beta/2$ and $5\beta/2$ separates the target's larger AC Stark phase from the background phases for QSP.

\subsection{Phase statistics at large detuning and long sensing times}
\label{subsec:sm_long_time}

The phase in Eq.~\eqref{eq:sm_phase_kernel} is quadratic in the
random mode amplitudes. 
With $\E[A_j^2]=2$,
$\E[A_j^4]=8$, 
and the independence of $\varphi_i$, 
we obtain
\begin{align}
 \E[A_jA_l e^{i(\varphi_l-\varphi_j)}]
 &=2\delta_{jl},\nonumber\\
 \E[A_jA_lA_mA_n
 e^{i(\varphi_l-\varphi_j+\varphi_n-\varphi_m)}]
 &=4(\delta_{jl}\delta_{mn}+\delta_{jn}\delta_{lm}),
 \label{eq:sm_mode_moments}
\end{align}
where $\delta_{jl}$ is the Kronecker delta. 
Applying these identities to the phase kernel yields its finite-time mean and variance,
\begin{align}
 \E[\theta_k^{(2)}]
 &= \frac{B_{\rm rms}^2}{2}\sum_j f(v_j)\Delta v\,
 \operatorname{Im}\mathcal K_{jj},
 \label{eq:sm_finite_means}\\
 \operatorname{Var} \theta_k^{(2)}
 &= \frac{B_{\rm rms}^4}{16}\sum_{j,l}f(v_j)f(v_l)(\Delta v)^2
 |\mathcal K_{jl}-\mathcal K_{lj}^*|^2.
 \label{eq:sm_finite_variance}
\end{align}
Here, the diagonal mode pairings, 
$\operatorname{Im}\mathcal K_{jj}$, 
determine the mean, 
whereas the variance also receives contributions from crossed pairings between different modes through $\mathcal K_{jl}-\mathcal K_{lj}^*$.
Both of the terms can therefore be expressed in terms of the Hermitian phase kernel
$\mathcal H_{jl}=(\mathcal K_{jl}-\mathcal K_{lj}^*)/(2i)$.
Its diagonal elements $\mathcal H_{jj}=\operatorname{Im}\mathcal K_{jj}$ determine the mean, 
while $|\mathcal H_{jl}|^2$ weights each mode pairing in the variance. 
To evaluate this kernel for smooth,  
off-resonant driving, 
we first integrate the inner integral in Eq.~\eqref{eq:smfilters} by parts. 
Using $\chi(0)=0$, we obtain
\begin{equation}
 \int_0^t  {\rm d} t'\, \chi(t') e^{-i\Delta_{k,l} t'}
 =\frac{i \chi(t) e^{-i\Delta_{k,l}t}}{\Delta_{k,l}}
 -\frac{i}{\Delta_{k,l}} \int_0^t {\rm d} t'\, \chi'(t') 
 e^{-i\Delta_{k,l} t'}.
\end{equation}
For a fixed smooth envelope $\chi(t)=w(t/T)$ and $|\Delta|T\gg1$, 
the second term involving the derivative of $\chi$ is strongly suppressed.
Keeping only the leading boundary contribution
gives
$\mathcal K_{jl}\simeq
 i\int_0^T{\rm d} t\, \chi(t)^2e^{i(q_l-q_j)t}/\Delta_{k,l}$.
Combining this expression with the corresponding $\mathcal K_{lj}^*$ term yields
\begin{equation}
 \mathcal H_{jl} \simeq \frac12
 \left(\frac{1}{\delta_k-q_j}
 + \frac{1}{\delta_k-q_l} \right)
 \int_0^T {\rm d} t\, \chi(t)^2 e^{i(q_l-q_j)t}.
 \label{eq:sm_dispersive_kernel}
\end{equation}
This approximation preserves the individual detuning of each mode, 
while the factor $e^{i(q_l-q_j)t}$ captures the interference between each pair of modes over the sensing window.

We now assume that the detuning is sufficiently large, i.e., $q_{\max}\ll|\delta_k|$, so both inverse detunings in Eq.~\eqref{eq:sm_dispersive_kernel} reduce to $1/\delta_k$. 
Substituting into Eq.~\eqref{eq:sm_phase_kernel} gives
\begin{equation}
 \theta_k^{(2)}
 \simeq
 \frac{B_{\rm rms}^2}{4\delta_k}
 \int_{t_0}^{t_0+T} {\rm d} t \,\chi(t-t_0)^2
 \left|\sum_j A_j \sqrt{f(v_j) \Delta v}\,
 e^{-i(q_jt+\varphi_j)}\right|^2.
 \label{eq:sm_local_intensity}
\end{equation}
The phase variance in the small-$\Delta v$ limit is therefore
\begin{equation}
 \operatorname{Var}\theta_k^{(2)}
 \simeq \frac{B_{\rm rms}^4}{4\delta_k^2}
 \int_0^T {\rm d} t \int_0^T {\rm d} t' \,
 \chi(t)^2\chi(t')^2
 \left|\int_0^{v_{\max}}{\rm d} v\,f(v)
   e^{-im_av^2(t-t')/2}\right|^2.
 \label{eq:sm_intensity_variance}
\end{equation}
Both times $t$ and $t'$ lie in $[0,T]$,
so their difference $\tau=t-t'$ spans $[-T,T]$. 
We now consider the long duration limit, $T\gg\tau_{\rm DM}$. 
The variance is dominated by time separations over which the DM intensity remains correlated, and 
for a fixed smooth function $\chi(t)=w(t/T)$, 
$\chi(t)$ varies negligibly over this scale. 
We can therefore approximate $\chi(t)^2\chi(t-\tau)^2\simeq\chi(t)^4$, 
so that the leading time integral becomes $\int_0^T {\rm d} t\, \chi(t)^4=T\eta_4$. 
After changing variables to the time difference $\tau$, 
the remaining integral in this limit is
\begin{equation}
 \int_{-T}^{T} {\rm d} \tau
 \left|\int_0^{v_{\max}} {\rm d} v 
 \,f(v) e^{-im_av^2\tau/2} \right|^2
 \simeq
 2\pi \int_0^{v_{\max}} {\rm d} v\,
 \frac{f(v)^2}{m_av}.
 \label{eq:sm_correlation_integral}
\end{equation}
Restoring the physical duration $T$, 
the mean acquires the weight factor,
$\int_0^T{\rm d} t\, \chi(t)^2=T\eta_2$, 
while the long duration variance follows from
Eq.~\eqref{eq:sm_correlation_integral}.
Combining these results, 
we obtain the long duration mean and standard deviation,
\begin{equation}
 \overline\theta_k
 \simeq\frac{B_{\rm rms}^2 T \eta_2}{2 \delta_k},
 \qquad
 \sigma_{\theta,k}
 \simeq\frac{B_{\rm rms}^2}{2|\delta_k|}
 \left[2\pi T \eta_4
    \int_0^{v_{\max}} {\rm d} v\, \frac{f(v)^2}{m_a v}
\right]^{1/2}.
 \label{eq:sm_longtime_res}
\end{equation}
Their ratio is
\begin{equation}
 \frac{\sigma_{\theta,k}}{|\overline\theta_k|}
 \simeq\sqrt{\frac{\tau_{{\rm eff},k}}{T}},
 \label{eq:sm_long_time_scaling}
\end{equation}
where the effective correlation time is
\begin{equation}
 \tau_{{\rm eff},k}
 \equiv
 \frac{2\pi\eta_4}{\eta_2^2}
 \int_0^{v_{\max}} {\rm d} v\, 
 \frac{f(v)^2}{m_av}.
 \label{eq:sm_effective_averaging_time}
\end{equation}
These results show that, at fixed drive strength,
the mean phase grows linearly with $T$, 
while its relative standard deviation decreases as $T^{-1/2}$. 
At fixed mean phase, increasing the sensing duration allows a weaker effective DM–sensor coupling, 
$B_{\rm rms}\propto T^{-1/2}$, 
and simultaneously reduces the absolute phase fluctuation as $\sigma_{\theta,k}\propto T^{-1/2}$. 
A longer sensing duration therefore produces a more stable accumulated phase.

\subsection{Direct numerical evolution of the ESU}
\label{subsec:sm-direct}

For a fixed DM configuration $\xi$, 
the ESU evolution can be constructed numerically from the two-level RWA Hamiltonian. 
For a given branch $k$ and sensing interval $[t_0,t_0+T]$, we write
\begin{equation}
 \widetilde H_k^{(\xi)}(t)=\frac12
 \begin{pmatrix}
 \delta_k&h(t)\\
 h(t)^*&-\delta_k
 \end{pmatrix},
 \qquad
 h(t) = B_{\rm rms} \chi(t-t_0)
 \sum_j A_j \sqrt{f(v_j) \Delta v}\,
 e^{-i(q_j t + \varphi_j)}.
\end{equation}
Divide the sensing interval into $N_t$ steps of duration $\Delta t=T/N_t$.
At the midpoint of each step, $t_n=t_0+(n+1/2)\Delta t$, 
set $h_n=h(t_n)$ and
$\Omega_n=\sqrt{\delta_k^2+|h_n|^2}$, 
for $n=0,\ldots,N_t-1$.
Since $\widetilde H_k^{(\xi)}(t_n)^2=\Omega_n^2I_S/4$, 
treating the Hamiltonian as constant over each step gives an evolution operator for a duration $\Delta t$
\begin{equation}
 M_n 
 = e^{-i\widetilde H_k^{(\xi)}(t_n) \Delta t}
 =
 \cos(\Omega_n \Delta t/2)I_S
 -
 i \frac{\sin(\Omega_n\Delta t/2)}{\Omega_n}
 \begin{pmatrix}
 \delta_k & h_n\\
 h_n^* & -\delta_k
 \end{pmatrix}.
 \label{eq:sm_direct_step}
\end{equation}
Let $\widetilde U^{[n]}$ denote the propagator accumulated after $n$ steps. 
Starting from the identity, 
the ESU evolution is composed as
\begin{equation}
 \widetilde U^{[0]} = I_S,
 \qquad
 \widetilde U^{[n+1]} = 
 M_n\widetilde U^{[n]},\qquad
 \widetilde U^{[N_t]} = 
 M_{N_t-1}\cdots M_1M_0.
\end{equation}
After removing the free precession proportional to $\delta_k$,
one can obtain the ESU matrix in Eq.~\eqref{eq:esu_components} numerically.
The signal-induced phase can then be extracted from its diagonal element as
$\theta_k  = - \arg \zeta_k$.
For $T \gg \tau_{\rm DM}$, the intensity integral in Eq.~\eqref{eq:sm_local_intensity} contains contributions from many uncorrelated time intervals. 
A central-limit argument therefore predicts an approximately Gaussian phase distribution, 
with the mean and variance calculated in Eq.~\eqref{eq:sm_longtime_res}.

\begin{figure}[!t]
 \centering
 \includegraphics[width=0.72\linewidth]{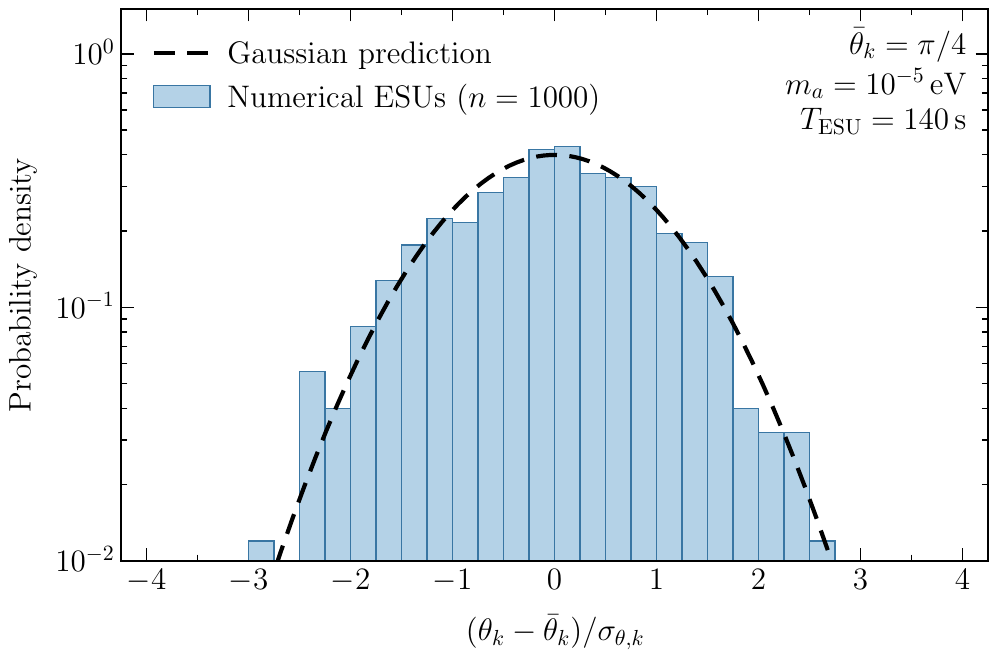}
 \caption{Distribution of the accumulated ESU phase $\theta_k$ for 1000 realizations of the stochastic DM waveform, 
 with $T=140\,\mathrm{s}$, $\overline\theta_k=\pi/4$, and $m_a=10^{-5}\,\mathrm{eV}$. 
 The histogram shows $(\theta_k-\overline\theta_k)/\sigma_{\theta,k}$.
 The dashed curve shows the Gaussian prediction for $\theta_k$, 
 with mean $\overline\theta_k=\pi/4$ and variance given by Eq.~\eqref{eq:sm_long_time_scaling}. We see that in the large-$T$ regime, the true $\theta_k$ distribution is well approximated by a Gaussian distribution.}
 \label{fig:sm_esu_phase_gaussian}
\end{figure}

We test this prediction using $10^3$ independent ESUs with different realizations of the random initial axion configuration $\xi$. 
Figure.~\ref{fig:sm_esu_phase_gaussian} shows that the simulated $\theta_k$ distribution agrees with the Gaussian prediction within the sampling uncertainty, 
supporting the Gaussian approximation in this long-duration regime.

\section{Quantum signal processing and frequency search}
\label{sm:qsp}

\subsection{QSP for identical ESUs}
\label{subsec:sign-QSP}

QSP converts the signal-dependent ESU phase into the sign required for oracle. 
For a single marked signal bin label $\ket{k_*}$, 
the ideal oracle acts as $\mathcal O_{\rm id}\ket k\ket\psi_S =(-1)^{\delta_{k,k_*}}\ket k\ket\psi_S$ for any sensor state $\ket\psi_S$, 
where $\delta_{k,k_*}$ is one for the marked label and zero otherwise. 
QSP produces this response by
interleaving ESU with programmed phase
rotations~\cite{Allen:2025hdx,Low:2016sck,Martyn:2021eaf}.
For a fixed phase $\theta$, each ESU implements $U(\theta) = e^{-i\theta Z_S}$.
With $L$ ESUs and $L+1$ rotations
$R_X(\phi)=e^{-i\phi X_S}$, the QSP unitary on each branch is
\begin{equation}
 Q_L(\theta)
 =
 R_X(\phi_L) U(\theta)\cdots
 R_X(\phi_1) U(\theta) R_X(\phi_0).
 \label{eq:sm_qsp_static}
\end{equation}
The programmed angles $\phi_\ell$ control interference between the $Z_S$ and $X_S$ rotations.
In this paper, 
we adopt a commonly used symmetric sequence of QSP control angles satisfying
\begin{equation}
 \phi_L
 =
 \phi_0 - \frac{\pi}{2},
 \qquad
 \phi_\ell
 =
 \phi_{L-\ell}\quad(1\leq \ell\leq L-1).
 \label{eq:sm_qsp_symmetry}
\end{equation}
For identical ESUs, 
this choice makes the diagonal entries of $Q_L(\theta)$ real.

The fixed phase QSP response, 
$P(x) \equiv \operatorname{Re} \bra{0} Q_L(\theta) \ket{0} $,
with $x=\cos\theta$, is a polynomial of degree at most $L$,
with $|P(x)|\leq1$ on $[-1,1]$. 
We choose $P$ to approximate the target sign $\varsigma(\theta)$, 
which is required to be $-1$ on the signal phase interval $\mathcal W_{\rm sig}$ and $+1$ on the background phase interval $\mathcal W_{\rm bg}$.
The QSP approximation error is the maximum scalar error over these intervals,
\begin{equation}
 \epsilon_{\rm QSP}
 =
 \max_{\theta \in \mathcal W_{\rm sig} \cup \mathcal W_{\rm bg}}
 \left[1 - \varsigma(\theta) P(\cos\theta) \right].
 \label{eq:sm_qsp_scalar_error}
\end{equation}
This scalar error bounds the deviation of the QSP unitary from its target oracle sign.
Since every QSP block belongs to $SU(2)$, 
it satisfies
$Q+Q^\dagger=2\operatorname{Re}(Q_{00})I_S$,
where $Q_{00}=\bra0Q\ket0$. 
Therefore,
\begin{equation}
 \begin{aligned}
 \|Q_L - \varsigma I_S \|
 = \sqrt{2 [1-\varsigma \operatorname{Re}(Q_{00})]}
 \leq \sqrt{2\epsilon_{\rm QSP}}.
 \end{aligned}
 \label{eq:sm_qsp_operator_error}
\end{equation}
Here, the operator norm is the largest singular value of the matrix difference. 
This error bound holds for any sensor input and therefore controls both population transfer, 
represented by the off-diagonal entries of $Q$,
and deviations from the target sign $\varsigma$.

The gap between $\mathcal W_{\rm sig}$ and $\mathcal W_{\rm bg}$ sets the required sequence length.
A narrower gap admits more phases but requires a longer sequence for the same accuracy. 
For separation $\Delta_x$ in $x=\cos\theta$,
bounded polynomial constructions achieve error $\epsilon_{\rm QSP}$ with $L = O[\Delta_x^{-1} \log(1/\epsilon_{\rm QSP})]$~\cite{Allen:2025hdx}.
Thus, $\epsilon_{\rm QSP}$ decreases exponentially in $L\Delta_x$ when an appropriate set of $\phi_\ell$ is chosen. 
For the $L=40$ sequence used in this paper, 
we choose the target signal phase range $\theta_k \in \mathcal W_{\rm sig} = [\pi/4,3\pi/4]$ and the background phase range with $|\theta_k| \leq 3\pi/20$. 
In terms of $x=\cos\theta_k$, 
the corresponding even polynomial separates the regions $|x|\leq\cos(\pi/4)$ and $|x|\geq\cos(3\pi/20)$, with a gap
$\Delta_x=\cos(3\pi/20)-\cos(\pi/4) \approx 0.184$.
Numerical checks give maximum error $\epsilon_{\rm QSP} \approx 1.36\times10^{-3}$ in the monochromatic limit, where all ESUs are identical.
The error bound does not apply to the gap $3\pi/20 \leq \theta_k \leq \pi/4$ where $P(x)$ is left unconstrained. 

For the frequency bins other than the signal bin, the larger detuning strongly suppresses the accumulated phase.
We therefore approximate the background sensing phase as zero. 
The corresponding ESUs reduce to identity operations 
while the programmed rotations remain, generating a QSP unitary on background bins as
\begin{equation}
 Q_b = 
 \exp\!\left[ - i
 \left(\sum_{\ell=0}^{L}\phi_\ell\right)X_S\right].
 \label{eq:sm_background}
\end{equation}
For the $L = 40$ sequence used in this paper, 
numerical evaluation gives $Q_b \approx 0.9986 I_S + 0.0520 i X_S$. 

\subsection{DM-induced phase fluctuations and stochastic errors}
\label{subsec:ESU-fluctuation}

For stochastic DM, the accumulated phase fluctuates around its mean value from one ESU to another. 
Even when these fluctuations remain within $\mathcal W_{\rm sig}$ or $\mathcal W_{\rm bg}$, 
variations among ESUs can alter the interference within a QSP sequence.

To isolate the effect of phase fluctuations, 
we fix the QSP angles and choose a mean phase $\bar\theta$ within the $\mathcal W_{\rm sig}$.
Let $Q_L^{(\xi)}$ denote the oracle matrix in Eq.~\eqref{eq:sm_qsp_static}, 
with the phase of ESU $\ell$ replaced by $\theta_\ell^{(\xi)}
=
\bar\theta + \delta\theta_\ell^{(\xi)}$,
the QSP approximation error at this mean phase is
$\epsilon_0(\bar \theta)
=1-\varsigma\operatorname{Re}\bra0Q_L(
\bar \theta)\ket0$, which is bounded by $\epsilon_{\rm QSP}$. 
As $\theta_\ell^{(\xi)}$ fluctuates around $\bar{\theta}$,
the expected error is
\begin{equation}
 \bar \epsilon(\bar\theta)
 = \mathbb E\!
 \left[
 1-\varsigma \operatorname{Re}
 \bra0 Q_L^{(\xi)} \ket0 \right].
 \label{eq:sm_qsp_fluctuation}
\end{equation}
This quantity is evaluated at the chosen mean phase rather than maximized over all possible $\theta_\ell^{(\xi)}$.
Subtracting the baseline $\epsilon_0(\bar{\theta})$ isolates the DM-induced stochastic error,
$\Delta
\epsilon_{\rm fluc}=
\overline \epsilon - \epsilon_0$.

Because $\delta \theta$ are approximately uncorrelated
between ESUs as the sensing time $T \gg \tau_{{\rm eff},k}$, the leading perturbations to the oracle accumulate as a random walk. 
Therefore, the mean squared operator distance from $Q_L^{(\xi)}$ to $Q_L(\bar\theta)$
grows in proportion to the number of ESUs.
For a fixed single-ESU variance $\sigma_\theta^2$, 
the stochastic error scales as $\Delta\epsilon_{\rm fluc}\propto L\sigma_\theta^2$.
Hence, we note a trade-off as we change $L$. 
While the QSP approximation error is suppressed as $L$ increases, the per-oracle stochastic error and timing complexity increases.

\begin{figure}[!t]
 \centering
 \includegraphics[width=0.96\textwidth]{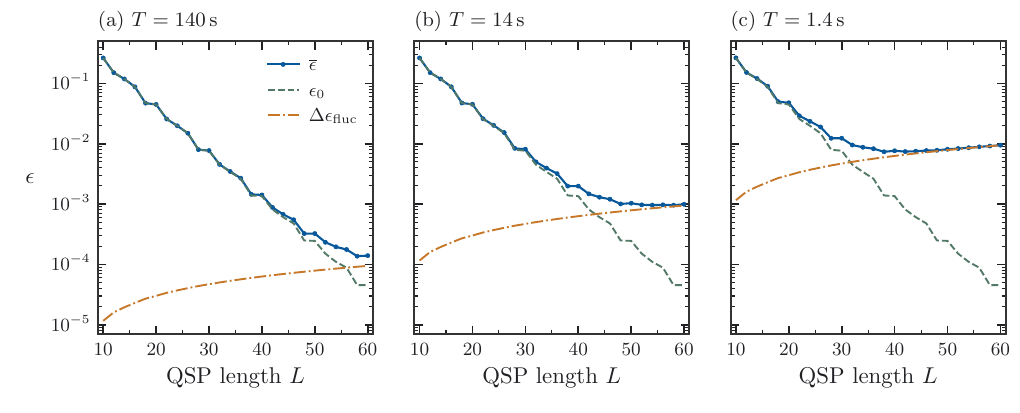}
 \caption{Scalar QSP errors at fixed $\bar \theta = \pi/4$: the total
 $\overline\epsilon$, 
 the QSP approximation error $\epsilon_0$, 
 and stochastic error $\Delta\epsilon_{\rm fluc}$. 
 We fix the ESU sensing time $T$ in each panel so that the standard deviation of the sensed phase $\sigma_{\theta,k}$ is fixed.
 These $\sigma_{\theta,k}$ are $0.0018$, $0.0056$, and $0.018$ for
 $T=140$, $14$, and $1.4\,\mathrm{s}$, respectively.
 The coupling strength $B_{\rm rms}$ is adjusted between panels so that the expected sensed phase $\bar{\theta}$ remains the same for a fair comparison.}
 \label{fig:sm_qsp_noise}
\end{figure}

At $m_a=10^{-5}\,\mathrm{eV}$, $\bar\theta=\pi/4$, 
$T=140\,\mathrm{s}$, and $L=40$, 
the standard deviation of the sensed phase is $\sigma_\theta \approx 1.777\times10^{-3}$. 
Direct evaluation at this reference phase gives $\epsilon_0(\pi/4) \approx 1.355\times10^{-3}$, 
close to the maximum scalar error $\epsilon_{\rm QSP}$ for this sequence.
The stochastic DM signal increases the error by $\Delta\epsilon_{\rm fluc} \approx 6.31\times10^{-5}$, 
yielding $\bar\epsilon(\pi/4) \approx 1.42\times10^{-3}$.
Figure~\ref{fig:sm_qsp_noise} compares these contributions as $L$ varies.
A longer sequence can reduce the QSP approximation error with appropriate QSP control phases $\phi_\ell$.
However, 
it also introduces more stochastic errors and increases the sensing time per oracle to $LT$. 
The competition between these effects produces an optimal sequence length $L$ that minimizes the total QSP error, as shown in the figure. 
At a fixed mean phase, 
shorter sensing time $T$ produces larger stochastic errors. Therefore, increasing $T$ improves the QSP accuracy as discussed in the main text.

Note that these single-oracle errors are distinct from the final search failure probability. 
The Grover search success probability depends on how errors from successive oracle calls accumulate, as discussed below.

\subsection{Search on the register space}
\label{subsec:Grover-search}

To perform the Grover search,
we initialize the index register to a uniform superposition $\ket s=N^{-1/2}\sum_{k=1}^N \ket k$ over the $N$ frequency bin labels and the sensor in $\ket0_S$. 
Each oracle call is followed by a register reflection $R_s=2\ket s\bra s-I_Q$ so that the full unitary is $R_s \otimes I_S$.
The index $r=0,\ldots,G$ counts the number of completed Grover iterations,
with $r=0$ corresponding to the initial state and $G$ to the final iteration before readout. 
We choose $G=\lfloor\pi\sqrt N/4\rfloor$,
which gives $G=117$ for $N=22500$.

Assuming that the background phase is negligible, all $N-1$ background labels undergo the same unitary $Q_b$ and remain equally weighted after each reflection. 
Their normalized superposition is
$\ket B = (N-1)^{-1/2} \sum_{k\ne k_*} \ket k$.
The signal and background components can then be described by the sensor state amplitude vectors,
$\boldsymbol{a}_r= (a_{r,0}, a_{r,1})^T$ and $\boldsymbol{b}_r = (b_{r,0},b_{r,1})^T$,
whose entries correspond to the component in the sensor basis states $\ket{0}_S$ and $\ket{1}_S$.
The two vectors are not individually normalized;
instead, their squared norms satisfy
$\|\boldsymbol a_r\|^2+\|\boldsymbol b_r\|^2=1$.
As an example, the initial sensor state is
\begin{equation}
 \boldsymbol a_0 = \frac1{\sqrt N}
 \begin{pmatrix}1\\0\end{pmatrix},
 \qquad
 \boldsymbol b_0 = \sqrt{\frac{N-1}{N}}
 \begin{pmatrix}1\\0\end{pmatrix}.
\end{equation}

On the $r$th iteration, the signal QSP unitary $Q_{s,r}$ acts on
$\boldsymbol{a}_{r-1}$ and the background QSP unitary $Q_b$ acts on $\boldsymbol{b}_{r-1}$.
The subsequent register reflection 
mixes these two vectors identically for both sensor components.
With $c_G = 1 - 2/N$ and $s_G = 2 \sqrt{N-1}/N$,
the resulting four-dimensional evolution is
\begin{equation}
 \begin{pmatrix}\boldsymbol a_r\\\boldsymbol b_r\end{pmatrix}=
 \begin{pmatrix}
 -c_G Q_{s,r} & s_G Q_b\\
 s_G Q_{s,r} & c_G Q_b
 \end{pmatrix}
 \begin{pmatrix}\boldsymbol a_{r-1}\\\boldsymbol b_{r-1}\end{pmatrix}.
 \label{eq:sm_search_recurrence}
\end{equation}
Each displayed matrix entry is a $2 \times 2$ unitary acting on the two-level sensor state, 
so the full matrix acts on four complex amplitudes.
At the final iteration $r = G$, 
the probability of measuring the signal label, summed over both sensor outcomes,  is
\begin{equation}
 P_{\rm bin}(\xi)
 =
 \boldsymbol a_G^\dagger
 \boldsymbol a_G
 =|a_{G,0}|^2 + |a_{G,1}|^2.
 \label{eq:grover_success}
\end{equation}

For an ideal oracle, $Q_{s,r}=-I_S$ and $Q_b=I_S$ so that the sensor returns to its initial state $\ket{0}_S$,  
and the repeated Grover iterations reduce to the ideal Grover rotation,
\begin{equation}
 P_{\rm id}
 =
 \sin^2[(2 G + 1)\alpha],\qquad
 \alpha=\arcsin(N^{-1/2}).
 \label{eq:sm_ideal_grover}
\end{equation}
For $N=22500, \, G=117$, 
this gives $P_{\rm id}=0.99998$.
With both the QSP approximation error and the stochastic error, however, the sensor state may not perfectly return to $\ket{0}_S$ and may obtain register-dependent components, weaken the Grover amplification to $\ket{k_*}$ from $\ket{s}$.
Eq.~\eqref{eq:sm_search_recurrence} then accounts for these $k$-dependent sensor states through all $G$ Grover iterations so that the dependence on $Q_{s,r}$ and $Q_b$ in the search's success probability given in Eq.~\eqref{eq:grover_success} can be tracked explicitly.

\subsection{Suppressing search error with an alternating-sign protocol}
\label{subsec:Alternating-QSP}
Following the procedure in Sec.~\ref{subsec:Grover-search}, we can estimate the search success probability $P_{\rm bin}(\xi)$ of a given QSP sequence.
For the $L=40$ benchmark, 
the marked bin has $\bar\epsilon(\pi/4)\simeq1.42\times10^{-3}$, 
corresponding to an RMS operator error of $\sqrt{2\bar\epsilon}\simeq0.053$. Taking this as a representative single-call error, 
a crude estimate that assumes coherent accumulation over $G = 117$ calls gives $G\sqrt{2\bar\epsilon}\gg1$. 
This suggests that oracle errors could substantially degrade the search.

Reducing the QSP approximation error would naively improve the search success probability. 
However, this generally requires longer sequences, 
which increase both the single-oracle sensing time and the stochastic error. 
Instead, we find that combining the original QSP oracle with a sign-flipped QSP oracle suppresses coherent accumulation of the static error at fixed $L$, 
making $L=40$ sufficient to find the marked bin in our benchmark.

To see how this works, 
we first take the monochromatic limit such that the ESUs experiences no stochastic error. 
Equation~\eqref{eq:sm_search_recurrence} then applies with fixed signal and background matrices $Q_s$ and
$Q_b$ for the sensor state. 
Let $G_+ = (R_s \otimes I_S)\mathcal O_+$ denote one search step, where
$\mathcal O_+ = \operatorname{diag}(Q_s,Q_b)$ is the oracle for signal and background.
Two calls in the conventional Grover search procedure with the same-sign protocol give
\begin{equation}
 G_+^2 =
 \begin{pmatrix}
 c_G^2 Q_s^2 + s_G^2 Q_b Q_s & c_G s_G (Q_b^2 - Q_s Q_b)\\
 c_G s_G (Q_b Q_s - Q_s^2) & s_G^2 Q_s Q_b + c_G^2 Q_b^2
 \end{pmatrix}.
 \label{eq:sm_same_sign_search}
\end{equation}
The diagonal terms contain $Q_s^2$ and $Q_b^2$ with coefficient
$c_G^2 = (1 - 2/N)^2 \simeq 1$. 
A fixed departure of these unitaries from the target signs ($-I_S$ and $I_S$) is therefore repeated. 
This error accumulates and affects the success probability even when the error of one call is small.

Suppose instead that the next oracle uses
$\mathcal O_- = \mathcal O_+^\dagger$, and let's define
$G_- = (R_s \otimes I_S)\mathcal O_-$. 
The two-step product becomes
\begin{equation}
 G_- G_+ =
 \begin{pmatrix}
 c_G^2 I_S + s_G^2 Q_b^\dagger Q_s &
 c_G s_G (I_S - Q_s^\dagger Q_b)\\
 c_G s_G (Q_b^\dagger Q_s - I_S) &
 s_G^2 Q_s^\dagger Q_b + c_G^2 I_S
 \end{pmatrix}.
 \label{eq:sm_alternating_search}
\end{equation}
Because $Q_{s/b}^\dagger Q_{s/b} = I_S$, the term $\propto c_G^2$ in $G_-G_+$ are now exactly proportionate to $I_S$ rather than $Q_{s/b}^2$. 
The remaining deviations from the ideal case carry a much smaller factor $s_G c_G$ or $s_G^2$, 
whereas the repeated same-sign call $G^2_+$ retains the terms $\sim \mathcal{O}(c_G^2)$, which is large.

For a quantitative comparison, let $G_{\rm id}$ be the ideal Grover iteration unitary with
$Q_s=-I_S$ and $Q_b=I_S$. 
Then, from Eq.~\eqref{eq:sm_qsp_operator_error}, we have
\begin{equation}
 \begin{aligned}
 \|G_+ G_+ - G_{\rm id}^2\|
 &\leq 2\sqrt{2\epsilon_{\rm QSP}},\\
 \|G_- G_+ - G_{\rm id}^2\|
 &=s_G\|Q_s+Q_b\|
 \leq 2s_G\sqrt{2\epsilon_{\rm QSP}}.
 \end{aligned}
 \label{eq:sm_alternating_bound}
\end{equation}
The first bound follows by adding the error bounds of the two steps. 
For the second bound, 
we multiply the matrix difference by $R_s \otimes I_S$, 
which leaves its norm unchanged. 
The diagonal blocks then vanish. 
Only the off-diagonal term of the matrix difference $s_G (I_S + Q_s^\dagger Q_b)$ and its adjoint remain. 
Since $Q_s$ is unitary, 
the norm of this matrix is $s_G\|Q_s+Q_b\|$.
Finally, $Q_s+Q_b = (Q_s+I_S) + (Q_b-I_S)$ gives the last inequality.
Thus, the maximum error in two calls is reduced by
$s_G\simeq2/\sqrt N$ in the monochromatic.
A similar argument also applies in the presence of small stochastic error.

\begin{figure}[!t]
 \centering
 \includegraphics[width=0.76\linewidth]{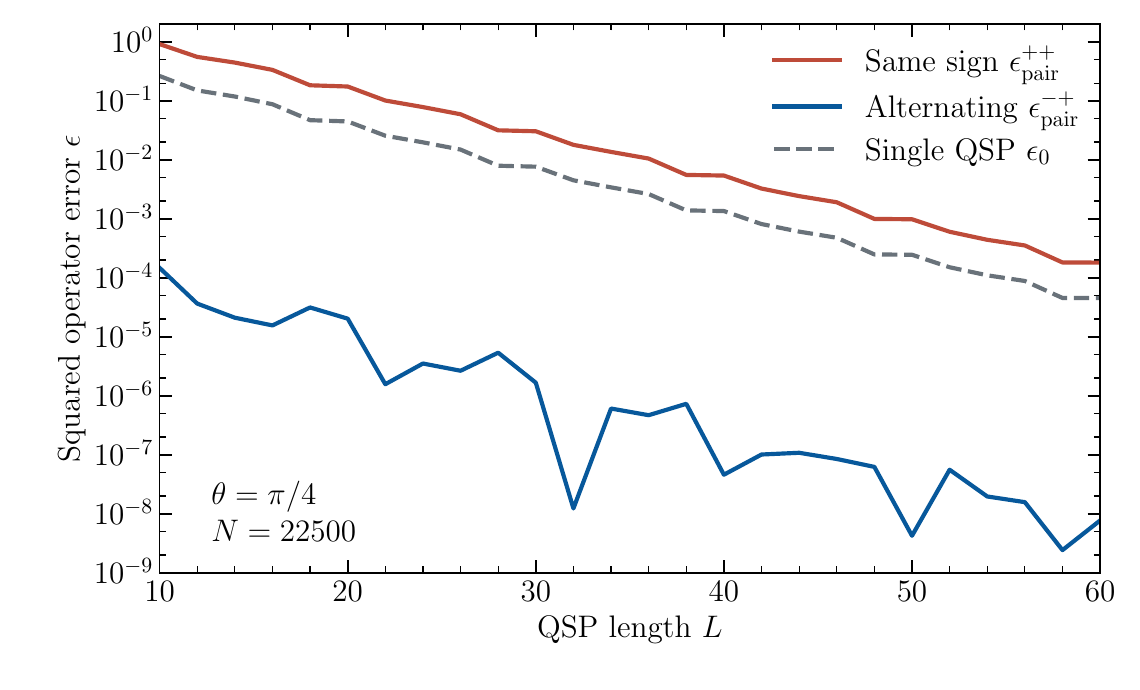}
 \caption{QSP approximation error suppression from alternating the signs of all QSP control angles,
 for identical diagonal ESUs with $\theta = \pi/4$, background oracle $Q_b$, and $N = 22500$.
 The gray dashed curve shows the single-QSP error $\epsilon_0$.
 The red and blue curves show the two-step squared errors $\epsilon_{\rm pair}^{++}$ and
 $\epsilon_{\rm pair}^{-+}$ for the same-sign and the alternating-sign protocols respectively. As discussed in the text, we find that $\epsilon_{\rm pair}^{-+} \simeq 4\epsilon_0/N \simeq \epsilon_{\rm pair}^{++}/N$ can suppresses the intrinsic QSP approximation error significantly.}
 \label{fig:sm_pair_error}
\end{figure}

Fig.~\ref{fig:sm_pair_error} showcases this error suppression. 
We define the squared errors over two Grover iterations as $\epsilon^{++}_{\rm pair} = \frac{1}{2} \|G_+ G_+ - G_{\rm id}^2\|^2$ and  $\epsilon^{-+}_{\rm pair} = \frac{1}{2} \|G_- G_+ - G_{\rm id}^2\|^2$. In the monochromatic limit with  background oracle $Q_b$, Eq.~\eqref{eq:sm_alternating_bound}
relates these errors to the single-QSP error $\epsilon_0 =\frac{1}{2} \|Q_s+I_S \|^2$,
with $\epsilon_{\rm pair}^{++} \simeq\left(1+c_G\right)^2 \epsilon_0 \simeq 4 \epsilon_0$ and $\epsilon_{\rm pair}^{-+}=s_G^2 \epsilon_0 \simeq 4 \epsilon_0/N \ll \epsilon^{++}_{\rm pair}$ for large number of frequency bins $N$.
The alternating-sign protocol therefore suppresses the static error without increasing the QSP sequence length.

To obtain $Q_{s/b}^\dagger$ using QSP, we simply flip the signs of all QSP angles in the symmetric sequence in Eq.~\eqref{eq:sm_qsp_symmetry}.  
In the monochromatic limit where all ESUs are identical, 
each factor in Eq.~\eqref{eq:sm_qsp_symmetry} is a symmetric matrix. 
The symmetry of the control angles, together with $R_X(\phi_L)=iX_SR_X(\phi_0)$, implies $Q_L^T=X_SQ_LX_S$. 
The two diagonal entries of $Q_L$ are therefore equal. 
Since $Q_L\in \mathrm{SU}(2)$, 
these entries are also complex conjugates and must consequently be real.
Let $Q_{+\boldsymbol\phi}$ and $Q_{-\boldsymbol\phi}$ denote the
original and sign-flipped sequence. Since
$Z_S R_X(\phi) Z_S = R_X(-\phi)$ and $Z_S U(\theta) Z_S = U(\theta)$, we have
\begin{equation}
 Q_{-\boldsymbol\phi}(\theta)
 = Z_S Q_{+\boldsymbol\phi}(\theta) Z_S
 = Q_{+\boldsymbol\phi}(\theta)^\dagger.
 \label{eq:sm_alternating_adjoint}
\end{equation}
The above relation holds exactly in the monochromatic limit, while the actual DM signal between ESUs introduce additional stochastic error. 
The sign flip therefore suppresses the QSP approximation error but not the stochastic error.

\begin{figure}[!t]
 \centering
 \includegraphics[width=0.96\textwidth]{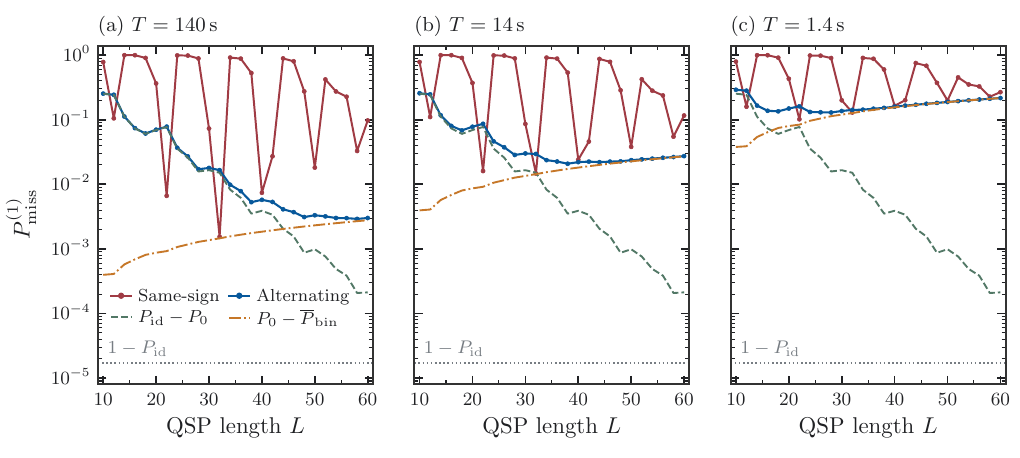}
 \caption{Single-search miss probability $P_{\rm miss}^{(1)}=1-\overline P_{\rm bin}$ for repeated QSP angles (same-sign, red) and alternating signs of all angles between oracle calls (alternating, blue), 
 with $T=140$, $14$, and $1.4\,\mathrm{s}$. 
 The green and orange curves show the alternating protocol’s miss probability due to QSP approximation error $P_{\rm id}-P_0$ and that due to the stochastic error $P_0-\overline P_{\rm bin}$, respectively,
 as decomposed in Eq.~\eqref{eq:sm_search_error_budget}. 
 The dotted line marks the finite-iteration residual $1-P_{\rm id}=1.70\times10^{-5}$, existing even for ideal Grover search algorithm. 
 All panels use a fixed mean phase for the marked bin $\bar{\theta}_{k_*} = \pi/4$ and $G=117$.}
 \label{fig:sm_grover_comparison}
\end{figure}

Figure~\ref{fig:sm_grover_comparison} compares the two protocols for $N=22500$, $G=117$, $m_a=10^{-5}\,\mathrm{eV}$, and $\beta=10^{-9}\,\mathrm{eV}$. 
Both use the same QSP sequence up to sign for each even length from $L=10$ to $60$. 
The ``same-sign" protocol repeats $\boldsymbol\phi$, 
whereas the ``alternating" protocol alternates between $\boldsymbol\phi$ and $-\boldsymbol\phi$.
For each ESU duration, $B_{\rm rms}$ is adjusted to keep the marked mean phase at $\pi/4$, so shorter ESU sensing time $T$ have larger phase variance.

Let $P_0$ denote the success probability obtained with
the ESU phase variance set to zero and
the ideal-$\pm I_S$-oracle probability $P_{\rm id}$ is given by
Eq.~\eqref{eq:sm_ideal_grover}. 
The single-search miss probability
can then be separated exactly as
\begin{equation}
 \begin{aligned}
 P_{\rm miss}^{(1)}
 &=1-\overline P_{\rm bin}\\
 &=(1-P_{\rm id})+(P_{\rm id}-P_0)
   +(P_0 - \overline P_{\rm bin}).
 \end{aligned}
 \label{eq:sm_search_error_budget}
\end{equation}
The first term is the finite-iteration error, exist even with ideal Grover search algorithm. 
The second measures the search loss from intrisic QSP approximation errors, 
and the third come from the the DM-induced stochastic noise.

The red curves show that the same-sign protocol can produce larger accumulated QSP approximation errors over the full search, 
even as increasing $L$ improves the single-call approximation. 
Alternating between the original QSP oracle and sign-flipped oracle suppresses these oscillations and gives more stable performance. 
The contribution from the QSP approximation error (green) generally decreases rapidly with $L$, initially reducing the total miss probability (blue).
However, the contribution from the stochastic error (orange) grows with $L$ and eventually limits this improvement.
Shorter ESUs average the field intensity less effectively, 
producing larger phase variance at the same mean phase. 
Consequently, the miss probability increases as $T$ decreases at fixed $L$.

\section{Search probability and sensitivity}
\label{sm:inference}

\subsection{Ensemble average and the 95\% sensitivity}
\label{subsec:sm-ensemble}

The ensemble-averaged success probability of a single search, $\overline P_{\rm bin}$, 
is defined in Eq.~\eqref{eq:PbinAver}. 
We evaluate this quantity numerically using Monte Carlo sampling. 
Specifically, we draw $M$ independent initial DM field configurations, $\xi_i=\{A_j,\varphi_j\}$.
such that we use the Monte Carlo method to calculate $\overline P_{\rm bin}$ numerically. 
Each configuration determines the DM waveform throughout one complete search. 
For each realization, we construct the ESU matrices as described in Sec.~\ref{subsec:sm-direct}, 
compose the QSP oracles and register reflections, 
and calculate the corresponding success probability $P_{\rm bin}(\xi_i)$ using Eq.~\eqref{eq:Pbinhistory}.
The ensemble average is then 
\begin{equation}
 \overline P_{\rm bin}
 =
 \frac{1}{M} 
 \sum_{i=1}^M P_{\rm bin}(\xi_i).
 \label{eq:sm_ensemble_average}
\end{equation}

\begin{figure}[!t]
 \centering
 \includegraphics[width=0.78\linewidth]{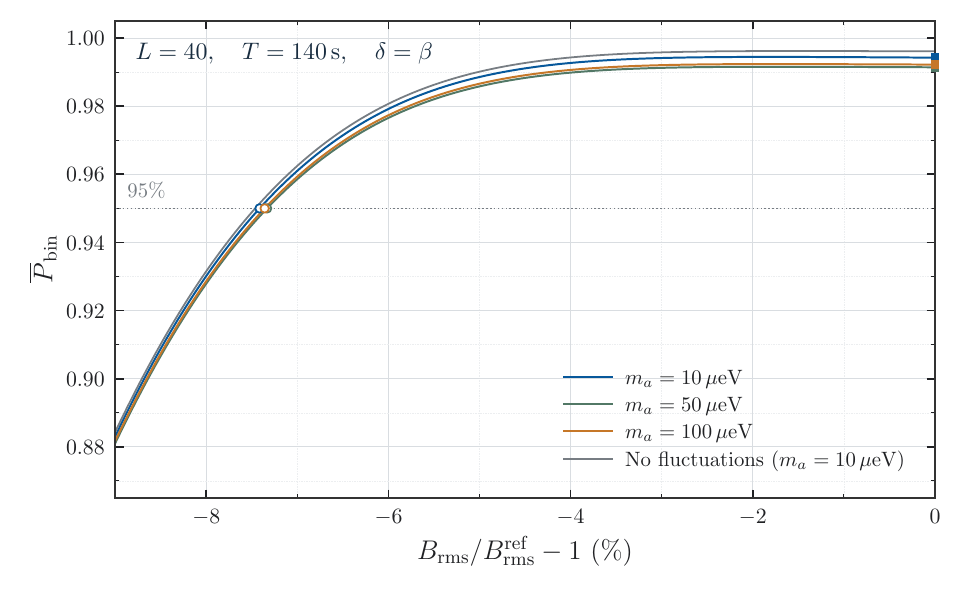}
 \caption{ Average search success probability near the $95\%$ threshold for representative DM masses. 
 The horizontal axis shows the fractional change in the field strength $B_{\rm rms}$ relative to the reference value $B_{\rm rms}^{\rm ref}$ in percentage. 
 The colored curves show the averaged success probability $\overline P_{\rm bin}$ of different stochastic DM signal with different mass for $T= 5 \times 10^5\tau_{\rm DM}(10\,\mu{\rm eV}) = 140\,$s, $L=40$, $N=22500$, $G=117$, and detuning $\delta$ equal to the bin width $\beta$. 
 The gray curve shows the corresponding result with a monochromatic signal while keeping the same finite length QSP sequence.}
 \label{fig:sm-boundary}
\end{figure}

To determine the sensitivity, 
we fix the DM mass $m_a$, 
the sensing duration $T$, 
and the QSP sequence length $L$. 
We then decrease the field strength from the reference value $B_{\rm rms}^{\rm ref}=B_{{\rm rms},\pi/4}$,
defined by requiring the mean accumulated phase in the signal bin to satisfy $\overline\theta_{k_*}=\pi/4$.
The first point at which the average success probability falls to $0.95$ defines the single-search sensitivity boundary,
\begin{equation}
 \overline P_{\rm bin}(B_{\rm rms}^{95\%}) 
 = 0.95,
 \qquad 
 P_{\rm miss}^{(1)}(B_{\rm rms}^{95\%})=0.05.
 \label{eq:sm_local_boundary}
\end{equation}
As shown in Fig.~\ref{fig:sm-boundary} and Table~\ref{tab:sm-boundaries}, 
the success probability at the reference field ranges from 99.1\% to 99.4\%. 
The 95\% boundary is reached after reducing the field strength by 7\% and lies close to the reference value.

\begin{table}[!t]
 \centering
 \begin{tabular}{rrrr}
 \hline\hline
 $m_a$ ($\mu\mathrm{eV}$)
 & $B_{{\rm rms},95}$ ($\mathrm{eV}$)
 & $g_{a\gamma\gamma,95}$ ($\mathrm{GeV}^{-1}$)
 & $\overline P_{\rm bin}$ at reference\\[1pt]
 \hline
 $10$ & $1.47\times10^{-13}$ & $3.18\times10^{-12}$ & $0.994$\\
 $20$ & $1.47\times10^{-13}$ & $4.48\times10^{-12}$ & $0.991$\\
 $50$ & $1.45\times10^{-13}$ & $7.01\times10^{-12}$ & $0.991$\\
 $100$ & $1.43\times10^{-13}$ & $9.71\times10^{-12}$ & $0.992$\\
 \hline\hline
 \end{tabular}
 \caption{Representative 95\% single-search sensitivities to the effective DM–sensor coupling strength $B_{\rm rms}$ and the axion–photon coupling $g_{a\gamma\gamma}$, 
 for fixed $T=5 \times 10^5\tau_{\rm DM}(10\,\mu{\rm eV}) = 140\,$s, $L=40$, $N=22500$, and $G=117$, with alternating QSP phases. 
 The final column gives the success probability at the reference field $B_{\rm rms}^{\rm ref}$, defined by $\overline\theta_{k_*}=\pi/4$.}
 \label{tab:sm-boundaries}
\end{table}

\subsection{Detection probability after repeated searches}
\label{subsec:sm-repetitions}

If a search misses the signal,
the full protocol can be repeated to increase the cumulative success probability. 
In each repetition, the control register and sensor are reinitialized, 
the complete QSP and Grover search protocol is executed, 
and the resulting singal bin is verified. 
Let $n$ denote the number of repeated searches. Assuming independent DM field realizations and the same average success probability for each search, 
the probabilities of missing the signal in all $n$ searches and detecting it at least once are
\begin{equation}
 P_{\rm miss}^{(n)}=(1-\overline P_{\rm bin})^n,
 \qquad
 P_{\rm det}^{(n)}=1-(1-\overline P_{\rm bin})^n.
 \label{eq:sm-repetitions}
\end{equation}

Table~\ref{tab:sm-durations} applies this repeated search strategy to shorter sensing durations. 
Each row is calibrated to the same mean phase, $\overline\theta_{k_*}=\pi/4$, 
so a shorter sensing duration requires a stronger field. 
However, shortening the sensing duration of each ESU increases the phase fluctuations and reduces the success probability of an individual search.
This loss can be compensated by repeating the complete search. 
We define $n_{95}$ as the smallest integer $n$ for which the cumulative detection probability satisfies $P_{\rm det}^{(n)}\geq0.95$.

\begin{table}[!t]
 \centering
 \begin{tabular}{rrrrrr}
 \hline\hline
 $T/\tauDM(m_a)$ & $\overline P_{\rm bin}$ & $n_{95}$
 & $P_{\rm det}^{(n_{95})}$ & $B_{{\rm rms},\pi/4}$ ($\mathrm{eV}$)
 & $n_{95} 2 G L T$ ($\mathrm h$)\\[1pt]
 \hline
 $5 \times 10^2$ & $0.372$ & $7$ & $0.961$ & $5.04\times10^{-12}$ & $2.55$\\
 $5 \times 10^3$ & $0.840$ & $2$ & $0.975$ & $1.59\times10^{-12}$ & $7.28$\\
 $5 \times 10^4$ & $0.978$ & $1$ & $0.978$ & $5.04\times10^{-13}$ & $36.4$\\
 $5 \times 10^5$ & $0.994$ & $1$ & $0.994$ & $1.59\times10^{-13}$ & $364$\\
 \hline\hline
 \end{tabular}
 \caption{Comparison of sensing durations and search repetitions at $m_a=10^{-5}\,\mathrm{eV}$.
 All rows use $L=40$, $N=22500$, and $G=117$.
 Each is independently calibrated to the same mean phase, $\overline\theta_{k_*}=\pi/4$.}
 \label{tab:sm-durations}
\end{table}

For $T=5 \times 10^3 \tau_{\rm DM}(10^{-5}\,\mathrm{eV}) = 1.4\,$s,
the single-search success probability at the reference field is $84.0\%$, 
so two independent searches yield a cumulative detection probability of $97.5\%$. 
We can therefore lower the field strength until the two search detection probability falls to $95\%$.
According to Eq.~\eqref{eq:sm-repetitions}, this requires $(1-\overline P_{\rm bin})^2=0.05$, 
or $\overline P_{\rm bin}=1-\sqrt{0.05} \simeq 0.776$ for each search.
At the mass $m_a=10^{-5}\,\mathrm{eV}$ used in Table~\ref{tab:sm-durations}, 
the corresponding effective DM–sensor coupling strength is
$ B_{{\rm rms},95}^{(n=2)} =
 1.45 \times 10^{-12}\,\mathrm{eV}$.
For comparison, 
a single search with $T=5 \times 10^5\tau_{\rm DM}(10^{-5}\,\mathrm{eV})$ reaches a $95\%$ success probability at $B_{{\rm rms},95} \simeq 1.47 \times10^{-13}\,\mathrm{eV}$
(Table~\ref{tab:sm-boundaries}). 
The two shorter searches require a total sensing time of $7.28\,\mathrm{h}$, 
compared with $364\,\mathrm{h}$ for the single longer search. 
This comparison illustrates the trade off between sensing time and sensitivity.
Repeated shorter searches achieve the desired detection probability in less time for stronger signals, 
whereas a longer search can detect weaker signals. 
We leave the optimization of this search strategy here to a future study.

\end{document}